\documentclass[11pt]{article}
\usepackage{amssymb}

\usepackage{graphics}
\usepackage{amsmath}
\usepackage{graphicx}
\usepackage{epsfig}
\usepackage{caption2}

\newtheorem{theorem}{Theorem}
\newtheorem{corollary}{Corollary}
\newtheorem{proposition}{Proposition}
\newtheorem{lemma}{Lemma}

\normalbaselineskip=\baselineskip

\renewcommand{\captionlabeldelim}{.}
\begin{document}

\title{Networked control systems: a perspective from chaos}
\author{{Guofeng Zhang } \\
{\small Dept. of Mathematical and Statistical Sciences}\\
{\small University of Alberta}\\
{\small Edmonton, Alberta, Canada \ T6G 2G1 }\\
{\small Email: \texttt{gzhang@math.ualberta.ca}} \and {Tongwen Chen} \\
{\small Dept. of Electrical and Computer Engineering}\\
{\small University of Alberta}\\
{\small Edmonton, Alberta, Canada \ T6G 2V4}\\
{\small Email: \texttt{tchen@ece.ualberta.ca}}}
\date{}
\maketitle

\begin{abstract}
In this paper, a nonlinear system aiming at reducing the signal
transmission rate in a networked control system is constructed by
adding nonlinear constraints to a linear feedback control system.
Its stability is investigated in detail. It turns out that this
nonlinear system exhibits very interesting dynamical behaviors: in
addition to local stability, its trajectories may converge to a
non-origin equilibrium or be periodic or just be oscillatory.
Furthermore it exhibits sensitive dependence on initial conditions
--- a sign of chaos. Complicated bifurcation phenomena are
exhibited by this system. After that, control of the chaotic
system is discussed. All these are studied under scalar cases in
detail. Some difficulties involved in the study of this type of
systems are analyzed. Finally an example is employed to reveal the
effectiveness of the scheme in the framework of networked control
systems.

\end{abstract}


\newpage
\listoffigures

\listoftables

\newpage

\addtolength{\baselineskip}{0.4\baselineskip}

\section{Introduction}
\subsection{Limited information related control}
 In the past decade great interest has been devoted to the study of
\textit{limited information related} control problems. Limited
information related control is defined as follows: Given a
physical plant $G$ and a set of performance specifications such as
tracking, design a controller $C$ based on limited information
such that the resulting closed-loop system meets the prespecified
performance specifications. There are generally two sources of
limited information, one is signal quantization, and the other is
signal transmission through various networks.

In designing a digital control system, signal quantization induced
by signal converters such as A/D, D/A and computer finite
word-length limitation is unavoidable. To compensate this,
traditional design methods generally proceed like this: First
design a controller ignoring the effect of signal quantization,
then model it as external white noise and analyze its effect on
the designed system. If the performance is acceptable, it is okay;
otherwise, adjust controller parameters such as the sampling
frequency, or do redesign (including the choice of converters)
until satisfactory performance is obtained. Recently the following
problems have been asked:
\begin{enumerate}
    \item How to study the effect of signal quantization more rigorously? More precisely,
how will it genuinely affect the performance of the underlying
control system?
    \item If there are positive answers to the above question, can one design
    better controllers based on this knowledge?
\end{enumerate}

To address these two problems, stability, the fundamental
requirement of a control system, has been studied recently in
somewhat detail. Delchamps [1990] studied the problem of
stabilizing an {\em unstable} linear time-invariant discrete-time
system via state feedback where the state is quantized by an
arbitrarily given quantizer of {\it fixed} quantization
sensitivity. It turned out that there are no state feedback
strategies ensuring asymptotic stability of the closed-loop system
in the sense of Lyapunov. Instead, the resulting closed-loop
system behaves chaotically. Fagnani \& Zampieri [2003] continued
this research in the context of a linear discrete-time scalar
system. Based on the flow information provided by the system
invoked by quantization, stabilizing methods based on the Lyapunov
approach and chaotic dynamics of the system were discussed. Ishii
\& Francis [2003] studied the quadratic stabilization of an
unstable linear time-invariant continuous-time system by designing
a digital controller whose input was the quantized system state;
an upper bound of sampling periods was calculated geometrically
using state feedback for the system $G$ with a carefully designed
quantizer of fixed quantization sensitivity, by which the
trajectories of the closed-loop system would enter and stay in a
region of attraction around the origin.  Clearly in order to
achieve asymptotic stability, quantizers with {\em variable}
quantization sensitivities must be adopted. In Brockett \&
Liberzon [2000], for the system $G$, by choosing a quantizer $q$
with time-varying sensitivities, a linear time-invariant feedback
was designed to yield global asymptotic stability. This problem
was also studied in Elia \& Mitter [2001] for exponential
stability using logarithmic quantizers. In Nair \& Evans [2002],
exponential stabilization of the system $G$ with a quantizer is
studied under the framework of probability theory. More
interestingly, the simultaneous effect of sampling period $T$ and
quantization sensitivity was studied in Bamieh [2003], where it is
shown via simulation that system performance would become
\textit{unbounded} as $T\rightarrow 0$ if a quantizer of fixed
sensitivity was inserted into a control loop composed of a system
and an unstable controller. Therefore it is fair to say that the
problem---performance of quantized systems--- is quite complicated
as well as challenging. Much research is still required in this
area.

Another representation of limited information is signals suffering
from time-delays or even loss, which are ubiquitous in the
networked control systems (Wong \& Brockett [1997], Walsh
\textit{et al.} [2001], and Ray [1987]). The fast-developing
secure, high speed networks (Varaiya \& Walrand [1996] and
Peterson \& Davie [2000]) make control over networks possible.
Compared to the traditional point-to-point connection, the main
advantages of connecting various system components such as
processes, controllers, sensors and actuators via communication
networks are wire reduction, low cost and easy installation and
maintenance, etc. Thanks to these merits, networked control
systems have been built successfully in various fields such as
automobiles (Krtolica \textit{et al.} [1994], Ozguner \textit{et
al.} [1992]), aircrafts (Ray [1987] and Sparks [1997]), robotic
controls (Malinowshi \textit{et al.} [2001], Safaric \textit{et
al.} [1999]) and so on. In addition, in the field of distributed
control, networks may provide distributed subsystems with more
information so that performance can be improved (Ishii \& Francis
[2002]). However, networks inevitably introduce time delays and
packet dropouts due to network propagation, signal computation and
coding, congestion, etc., which lead to limited information for
the system to be controlled as well as the controller, thus
complicating the design of controllers and degrading the
performance of control systems or even destabilizing them (Zhang
\textit{et al.} [2001]). Therefore it is very desirable to reduce
time delays and packet dropouts when implementing a networked
control system. For the limitation of space, for now we will
concentrate on discussing a network protocol proposed by Walsh,
Beldiman, Bushnell, and Hong, \textit{et al.} (Walsh \textit{et
al.} [1999, 2001, 2002a, 2002b]) since our proposed one is in the
same spirit as theirs. For a more complete review on networked
control systems and more references, please refer to Zhang \& Chen
[2003].

\subsection{Network based control}
One effective way to avoid large time delays and high probability
of packet dropouts is by reducing network traffic. In a series of
papers published by Walsh, Beldiman, Bushnell, and Hong,
\textit{et al.} (Walsh \textit{et al.} [1999, 2001, 2002a,
2002b]), a network protocol called try-once-discard (TOD) is
proposed. In that scheme, there is a network along the route from
a MIMO plant to its controller. At each transmission time, each
sensor node calculates the importance of its current value by
comparing it with the latest one, the larger the difference is,
the more important the current value is, then the most important
one gets access to the network. For this scheme, based on the
Lyapunov method and the perturbation theory, a minimal time within
which there must have at least one network transmission to
guarantee stability of networked control systems is derived.

This network protocol, TOD, essentially belongs to the category of
dynamical schedulers. In comparison with static schedulers such as
token rings, it allocates network resources more effectively.
However, a supervisor computer, i.e., a central controller, is
required to compare those differences and decide which node should
get access to the network at each transmission time. It is
therefore complicated and possibly difficult to implement. In this
paper, we introduce another technique aiming at reducing network
traffic.

\subsection{A new networked control technique}
Consider the feedback system in Fig. 1,
\begin{figure}[tbh]
\epsfxsize=4in
\par
\epsfclipon
\par
\centerline{\epsffile{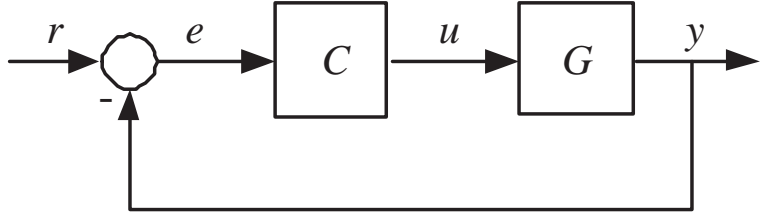}} \caption{A standard feedback
system}
\end{figure}
where $G$ is a discrete-time system of the form:
\begin{eqnarray}
x(k+1) &=&Ax(k)+Bu(k),  \label{sysG} \\
y(k) &=&Cx(k),  \notag
\end{eqnarray}
with the state $x\in \mathbb{R}^{n}$, the input $u\in
\mathbb{R}^{m}$, the output $y\in \mathbb{R}^{p}$ and the
reference input $r\in \mathbb{R}^{p}$ respectively; $C$ is a
stabilizing controller:
\begin{eqnarray}
x_{d}(k+1) &=&A_{d}x_{d}(k)+B_{d}e(k),  \label{conC} \\
u(k) &=&C_{d}x_{d}(k)+D_{d}e(k),  \notag \\
e\left( k\right) &=&r\left( k\right) -y\left( k\right) ,  \notag
\end{eqnarray}
with its state $x_{d}\in \mathbb{R}^{n_{c}}$. Let $\xi =\left[
\begin{array}{c}
x \\
x_{d}%
\end{array}
\right] $, then the closed-loop system from $r$ to $e$ can be modeled by
\begin{eqnarray}
\xi \left( k+1\right) &=&\left[
\begin{array}{cc}
A-BD_{d}C & BC_{d} \\
-B_{d}C & A_{d}%
\end{array}
\right] \xi \left( k\right) +\left[
\begin{array}{c}
BD_{d} \\
B_{d}%
\end{array}
\right] r(k),  \label{clsys1} \\
e(k) &=&\left[
\begin{array}{cc}
-C & 0%
\end{array}
\right] \xi \left( k\right) +r(k).  \notag
\end{eqnarray}

Now we add nonlinear constraints on both $u$ and $y$. Specifically, consider
the system in Fig. 2.
\begin{figure}[tbh]
\epsfxsize=4in
\par
\epsfclipon
\par
\centerline{\epsffile{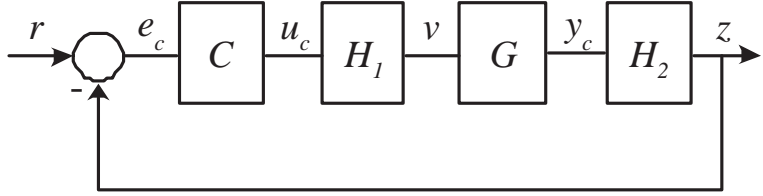}}
\caption{A constrained feedback system}
\end{figure}
The nonlinear constraint $H_{1}$ is defined as, for a given $\delta _{1}>0$,
let $v(-1)=0$, and for $k\geq0$,
\begin{equation}
v(k)=H_{1}\left( u_{c}\left( k\right) ,v(k-1)\right) =\left\{
\begin{array}{ll}
u_{c}(k), & \mbox{if~}\left\| u_{c}\left( k\right) -v\left( k-1\right)
\right\| _{\infty }>\delta _{1}, \\
v(k-1), & \mbox{otherwise.}%
\end{array}
\right.  \label{constraint1}
\end{equation}
Similarly $H_{2}$ is defined as, for a given $\delta _{2}>0$, let $z(-1)=0$,
and for $k\geq0$,
\begin{equation}
z(k)=H_{2}\left( y_{c}\left( k\right) ,z(k-1)\right) =\left\{
\begin{array}{ll}
y_{c}(k), & \mbox{if~}\left\| y_{c}\left( k\right) -z\left( k-1\right)
\right\| _{\infty }>\delta _{2}, \\
z(k-1), & \mbox{otherwise.}%
\end{array}
\right.  \label{constraint2}
\end{equation}
It can be shown that $\left\| H_{1}\right\| $, the induced norm of $H_{1}$,
equals $2$, so is $\left\| H_{2}\right\| $.

In a networked control system, there are normally computer networks along
the routes from the controller $C$ to the system $G$ and from $G$ to $C$.
These networks(usually shared by other clients) will introduce time delays
into the closed-loop system. It it quite appealing to compensate this
adverse effect. If we regard $H_{1}$ as a component of $C$ and $H_{2}$ of $G$%
, $G( \mbox{ resp. } C)$ contains previous version of $u_{c}( \mbox{ resp. }
y_{c}$), then there will have no signal transmission from $C$ to $G$ and(or)
from $G$ to $C$ if the inequalities in Eqs. (\ref{constraint1})-(\ref{constraint2}%
) are not satisfied, suggesting that we are reducing network
traffic. We expect this will benefit the overall system connected
by the common networks. One example will be given in Sec. 3 to
illustrate this point.

Similar work is done in Otanez \textit{et al.} [2003] where \textit{%
adjustable} deadbands are proposed to reduce network traffics. In that
formulation, the closed-loop system with deadbands is modeled as a perturbed
system, then its exponential stability follows that of the original system
[Khalil 1996]. The constraints proposed here are fixed ($\delta _{1}$ and $%
\delta _{2}$), we will see the stability of the system in Fig. 2
is quite complicated (e.g., only local stability can be obtained).
However, the advantage of fixed deadbands is that it will reduce
network traffic more effectively. Furthermore, the stability
region can be scaled as large as desired. This is one advantage of
our proposed scheme. Moreover, we find out that the system in Fig.
2 has rather complex dynamics --- it appears chaotic. As is known
chaotic behavior will in general provide more system dynamics,
i.e., more information of the underlying system, therefor we hope
we can achieve better control in the framework of Fig. 2.  We will
address this problem more rigorously in Sec. 2.2.

For the ``constrained'' system in Fig. 2, let $p$ denote the state
of the system $G$, and $p_{d\text{ }}$denote the state of the
controller $C$, then
\begin{eqnarray*}
p(k+1) &=&Ap(k)+Bv(k), \\
y_{c}(k) &=&Cp(k),
\end{eqnarray*}
and
\begin{eqnarray*}
p_{d}(k+1) &=&A_{d}p_{d}(k)+B_{d}e_{c}(k), \\
u_{c}(k) &=&C_{d}p_{d}(k)+D_{d}e_{c}(k), \\
e_{c}(k) &=&r(k)-z(k).
\end{eqnarray*}
Let $\eta =\left[
\begin{array}{c}
p \\
p_{d}%
\end{array}
\right] $, then the closed-loop system from $r$ to $e$ is
\begin{eqnarray}
\eta (k+1) &=&\left[
\begin{array}{cc}
A & 0 \\
0 & A_{d}%
\end{array}
\right] \eta (k)+\left[
\begin{array}{cc}
B & 0 \\
0 & B_{d}%
\end{array}
\right] \left[
\begin{array}{c}
v(k) \\
-z(k)%
\end{array}
\right] +\left[
\begin{array}{c}
0 \\
B_{d}%
\end{array}
\right] r(k),  \label{clsys2} \\
e_{c}(k) &=&\left[
\begin{array}{cc}
-C & 0%
\end{array}
\right] \eta \left( k\right) +r(k),  \notag
\end{eqnarray}
where $v$ and $z$ are given in Eqs.
(\ref{constraint1})-(\ref{constraint2}).

To test whether the scheme adopted here is useful in the framework
of networked control systems, we have to address at least the
following two concerns:

\begin{itemize}
\item The stability of the system in Fig. 2. Since stability is fundamental
to any control system, the first question about this system is its
stability. In this paper, the Lyapunov stability is studied in
detail:

\begin{enumerate}
\item Given that both $G$ and $C$ are stable, The system is \textit{locally}
exponentially stable (Lemma 1).

\item However, the behavior of the state trajectory ($p$, $p_{d}$), starting
outside the stability region, is hard to predict. A scalar case is
studied in detail to illustrate various dynamics the system can
exhibit (Sec. 2.1): Its trajectory may converge to an equilibrium
which is not necessarily the origin (Proposition 1, Corollary 1),
or be periodic (Theorem 3, Theorem 4), or aperiodic (Theorem 1,
Theorem 2), which can either be quasiperiodic or exhibit sensitive
dependence on initial conditions --- a sign of chaos, advocating
novel control method --- chaotic control.

\item For higher-order cases, a positively invariant set is constructed (Theorem 5).

\item Finally it is proved that the set of all initial points $\eta (0)$
whose closed-loop trajectories tend to an equilibrium as $k\rightarrow
\infty $ has Lebesgue measure zero if either $G$ or $C$ is unstable (Theorem
6).
\end{enumerate}

\item This research is mainly devoted to the study of networked control
systems (NCSs), hence it is natural and necessary to analyze its
effectiveness in the framework of networked control systems. An example is
used to illustrated the efficacy of our scheme (Sec. 3).
\end{itemize}

The outline of this paper as follows. Sec. 2 is devoted to the study of
stability. An example is
constructed to show the effectiveness of our scheme in Sec. 3. Some
concluding remarks are in Sec. 4.

\section{Stability}

In this section, we discuss the stability of the system in Eq.
(\ref{clsys2}). Firstly a sufficient condition ensuring local
exponential stability is derived. Secondly concentrated mainly on
 scalar cases, the intriguing behavior of the dynamics of the
system is studied in detail. It appears that the system behaves
chaotically. Finally it is proven that the Lebesgue measure of the
set of trajectories converging to a certain equilibrium is zero if
either the system $G$ or the controller $C$ is unstable.

Letting $r=0$, the system in Eq. (\ref{clsys2}) becomes
\begin{eqnarray}
\eta (k+1) &=&\left[
\begin{array}{cc}
A & 0 \\
0 & A_{d}%
\end{array}
\right] \eta (k)+\left[
\begin{array}{cc}
B & 0 \\
0 & B_{d}%
\end{array}
\right] \left[
\begin{array}{c}
v(k) \\
-z(k)%
\end{array}
\right] ,  \notag \\
\left[
\begin{array}{c}
u_{c}(k) \\
y_{c}\left( k\right)%
\end{array}
\right] &=&\left[
\begin{array}{cc}
0 & C_{d} \\
C & 0%
\end{array}
\right] \eta (k)+\left[
\begin{array}{cc}
0 & D_{d} \\
0 & 0%
\end{array}
\right] \left[
\begin{array}{c}
v(k) \\
-z(k)%
\end{array}
\right] ,  \label{switch} \\
\left[
\begin{array}{c}
v(k) \\
-z(k)%
\end{array}
\right] &=&\left[
\begin{array}{c}
H_{1}\left( u_{c}\left( k\right) ,v(k-1)\right) \\
-H_{2}\left( y_{c}\left( k\right) ,z(k-1)\right)%
\end{array}
\right], ~~ k\geq0.  \notag
\end{eqnarray}
Then, we have the following result regarding local stability.

\begin{lemma}
If both the system $G$ and the controller $C$ are stable, then the origin is
locally exponentially stable.
\end{lemma}

\noindent \textbf{Proof:} Define
\begin{equation*}
\tilde{A}=\left[
\begin{array}{cc}
A & 0 \\
0 & A_{d}%
\end{array}
\right] , ~~ \tilde{C}=\left[
\begin{array}{cc}
0 & C_{d} \\
C & 0%
\end{array}
\right] .
\end{equation*}
Since both $G$ and $C$ are stable, $\rho \left( \tilde{A}\right) <1$ where $%
\rho \left( M\right) $ is the spectral radius of a square matrix $M$. Then
for any given $\varepsilon >0$ satisfying $\rho \left( \tilde{A}\right)
+\varepsilon <1$, there exists a matrix norm $\left\| \cdot \right\| _{*}$
such that $\left\| \tilde{A}\right\| _{*}\leq \rho \left( \tilde{A}\right)
+\varepsilon $ [Huang, 1984]. Furthermore, this matrix norm satisfies $%
\left\| MN\right\| _{*}\leq \left\| M\right\| _{*}\left\| N\right\| _{*}$
for any two matrices $M$ and $N$ of dimension $n+n_{c}$. Therefore, for a
vector $x$ of dimension $n+n_{c}$, one can define a vector norm $\left|
x\right| _{*}$ such that $\left| Mx\right| _{*}\leq \left\| M\right\|
_{*}\left| x\right| _{*}$. One way to define such a norm is the following:
Let $\mathcal{O}$ denote the zero vector of dimension $n+n_{c}$, define
\begin{equation*}
\left| x\right| _{*}:=\left\| \left[
x,\underbrace{\mathcal{O},\cdots ,\mathcal{O}}_{n+n_{c}-1}\right]
\right\| _{*},
\end{equation*}
then
\begin{equation*}
\left| Mx\right| _{*}=\left\| \left[ Mx,\mathcal{O},\cdots ,\mathcal{O}%
\right] \right\| _{*}\leq \left\| M\right\| _{*}\left\| \left[ x,\mathcal{O}%
,\cdots ,\mathcal{O}\right] \right\| _{*}=\left\| M\right\| _{*}\left|
x\right| _{*}.
\end{equation*}
For a vector $\omega $ of dimension $\nu <n+n_{c}$, denote by $O$ the zero
vector of dimension $n+n_{c}-\nu $, define $\left| \omega \right|
_{*}:=\left| \left[
\begin{array}{ll}
\omega ^{^{\prime }} & O^{^{\prime }}%
\end{array}
\right] ^{^{\prime }}\right| _{*}$, then $\left| \cdot \right| _{*}$ is a
norm on the vector space $\mathbb{R}^{\nu \times 1}$. We treat a matrix of
dimension less than $n+n_{c}$ in the similar way.

Let $\left\| \cdot \right\| _{1}$ be the induced matrix norm of
the vector norm $\left\| \cdot \right\| _{\infty }$, then there
exist positive constants $c_{1}$ and $c_{2}$ such that
$c_{1}\left\| M\right\| _{*}\leq \left\| M\right\| _{1}\leq
c_{2}\left\| M\right\| _{*}$ for any matrix $M\in
\mathbb{R}^{n+n_{c}}$. Let $\delta :=\min \{\delta _{1}$, $\delta
_{2}\}$, then $\left\| M\right\| _{1}\leq \delta $ if $\left\|
M\right\| _{*}\leq
\delta /c_{2}$. Hence, in the sequel we concentrate on the matrix norm $%
\left\| \cdot \right\| _{*}$ and the upper bound $\delta /c_{2}$.
Now we are ready to derive the local stability of the system in
Eq.  (\ref{switch}). We claim that the stability region contains a
ball centered at the origin with radius
\begin{equation}
rd:=\min \left\{ \frac{\delta _{1}}{c_{2}\left\| \left[
\begin{array}{ll}
0 & C_{d}%
\end{array}
\right] \right\| _{*}},\frac{\delta _{2}}{c_{2}\left\| \left[
\begin{array}{ll}
C & 0%
\end{array}
\right] \right\| _{*}}\right\}  \label{radius}
\end{equation}
(denoted $\mathfrak{B}\left( 0,rd\right) $).

Suppose $\left| \eta (0)\right| _{\ast }\leq rd$, by Eq.
(\ref{switch}),
\begin{equation*}
\left| y_{c}(0)\right| _{\ast }\leq \left\| \left[
\begin{array}{ll}
C & 0%
\end{array}%
\right] \right\| _{\ast }\left| \eta (0)\right| _{\ast }\leq \frac{\delta
_{2}}{c_{2}},
\end{equation*}%
then
\begin{equation*}
\left\| y_{c}(0)\right\| _{\infty }\leq \delta _{2},
\end{equation*}%
hence
\begin{equation*}
z\left( 0\right) =H_{2}\left( y_{c}\left( 0\right) ,z(k-1)\right) =z\left(
-1\right) =0.
\end{equation*}%
Therefore
\begin{equation*}
\left| u_{c}(0)\right| _{\ast }\leq \left\| \left[
\begin{array}{ll}
0 & C_{d}%
\end{array}%
\right] \right\| _{\ast }\left| \eta (0)\right| _{\ast }\leq \frac{\delta
_{1}}{c_{2}},
\end{equation*}%
which means
\begin{equation*}
\left\| u_{c}(0)\right\| _{\infty }\leq \delta _{1},
\end{equation*}%
and
\begin{equation*}
v\left( 0\right) =H_{1}\left( u_{c}\left( 0\right) ,v(k-1)\right) =v\left(
-1\right) =0.
\end{equation*}%
Then
\begin{equation*}
\eta (1)=\tilde{A}\eta (0).
\end{equation*}%
Similarly,
\begin{eqnarray*}
\left| y_{c}(1)\right| _{\ast } &\leq &\left\| \left[
\begin{array}{ll}
C & 0%
\end{array}%
\right] \right\| _{\ast }\left| \eta (1)\right| _{\ast }=\left\| \left[
\begin{array}{ll}
C & 0%
\end{array}%
\right] \right\| _{\ast }\left\| \tilde{A}\right\| _{\ast }\left| \eta
(0)\right| _{\ast } \\
&\leq &\left\| \left[
\begin{array}{ll}
C & 0%
\end{array}%
\right] \right\| _{\ast }\left( \rho \left( \tilde{A}\right) +\varepsilon
\right) \left| \eta (0)\right| _{\ast }\leq \frac{\delta _{2}}{c_{2}},
\end{eqnarray*}%
\begin{equation*}
\left\| y_{c}(1)\right\| _{\infty }\leq \delta _{2},
\end{equation*}%
\begin{equation*}
H_{2}\left( y_{c}\left( 1\right) ,z(0)\right) =z\left( 0\right)
=0.
\end{equation*}%
Moreover,
\begin{eqnarray*}
\left| u_{c}(1)\right| _{\ast } &\leq &\left\| \left[
\begin{array}{ll}
0 & C_{d}%
\end{array}%
\right] \right\| _{\ast }\left| \eta (1)\right| _{\ast }=\left\| \left[
\begin{array}{ll}
0 & C_{d}%
\end{array}%
\right] \right\| _{\ast }\left\| \tilde{A}\right\| _{\ast }\left| \eta
(0)\right| _{\ast } \\
&\leq &\left\| \left[
\begin{array}{ll}
0 & C_{d}%
\end{array}%
\right] \right\| _{\ast}\left( \rho \left( \tilde{A}\right)
+\varepsilon \right) \left| \eta (0)\right| _{\ast }\leq \frac{\delta _{1}}{%
c_{2}},
\end{eqnarray*}%
which means
\begin{equation*}
\left\| u_{c}\left( 1\right) \right\| _{\infty }\leq \delta _{1},
\end{equation*}%
and
\begin{equation*}
v\left( 1\right) =H_{1}\left( u_{c}\left( 1\right) ,v(0)\right) =v\left(
0\right) =0.
\end{equation*}%
Then
\begin{equation*}
\eta (2)=\tilde{A}\eta (1)=\tilde{A}^{2}\eta (0)
\end{equation*}%
implying there is no updating for the inputs to $G$ and $C$. Following this
process, we see
\begin{equation*}
\eta (k)=\tilde{A}^{k}\eta (0)
\end{equation*}%
converges to zero as $k$ tends to $\infty $. $\hfill $ $\blacksquare $

\bigskip \noindent \textbf{Remark 1: } Though this system is locally
exponentially stable, it is hard to find the exact stability
region except for a scalar system controlled by a static feedback.
\ However, even in this scalar case, very complex dynamics can be
exposed by the system. This is the topic of the next subsection.

\subsection{Scalar case}

In this part, the definitions of such concepts as (positively)
invariant
sets, topological transitivity, structural stability, invariant sets and $%
\omega -$limit sets, etc., are adopted from Robinson [1995] or
Robinson [2004] unless otherwise specified.

To get a flavor of the complexity that the system in Fig. 2 may exhibit, we
first study a simple one-dimensional system:
\begin{eqnarray}
x(k+1) &=&ax(k)+bv(k),  \label{exap1} \\
u_{c}(k) &=&x(k),  \notag
\end{eqnarray}%
with $v(-1)\in \mathbb{R}$ without loss of generality, and for $k\geq 0$,
\begin{equation*}
v(k)=H_{1}\left( u_{c}\left( k\right) ,v(k-1)\right) =\left\{
\begin{array}{ll}
u_{c}(k), & \mbox{if~}\left| u_{c}\left( k\right) -v\left( k-1\right)
\right| >\delta _{1}, \\
v(k-1), & \mbox{else,}%
\end{array}%
\right.
\end{equation*}%
where $\delta _{1}=0.01$. The system in Eq. (\ref{exap1}) is a
static state feedback system with feedback gain equal to $1$. Note
that in this example
there is no constraint on the output of the system $G$. Now let $a=9/10$ and $%
b=-3/10$. By choosing different initial values $\left( v(-1),x(0)\right) $,
Figs. 3--4 are obtained.
\begin{figure}[tbh]
\epsfxsize=4in
\par
\epsfclipon
\par
\centerline{\epsffile{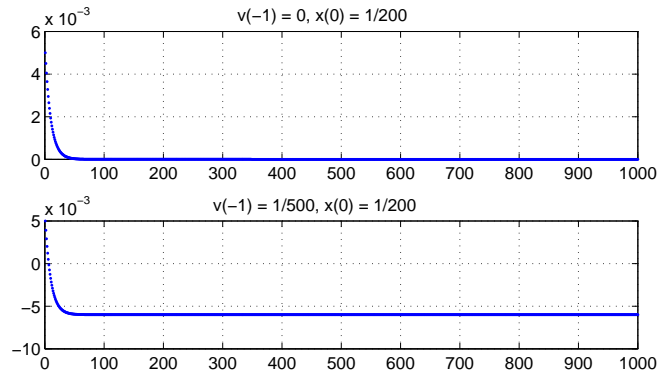}}
\caption{Two trajectories converging to two different fixed points}
\end{figure}
\begin{figure}[tbh]
\epsfxsize=4in
\par
\epsfclipon
\par
\centerline{\epsffile{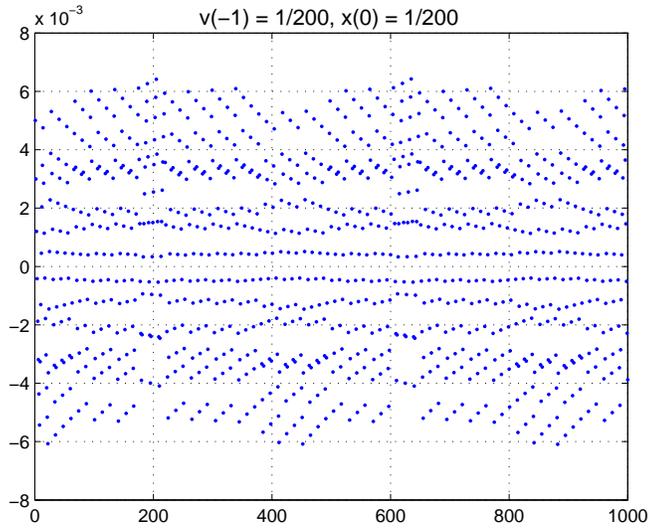}}
\caption{An aperiodic trajectory}
\end{figure}
In these two figures, the horizontal axis stands for the iteration
time $k$, and the vertical axis denotes the value of $x$. It is
clear from these two figures that different initial conditions
give rise to significantly different types of trajectories: the
first converging to the origin and the second converging to a
non-origin point and the last just oscillating. Furthermore, the
system in Eq. (\ref{exap1}) is actually able to exhibit
``chaotic'' behavior, i.e., sensitive dependence on initial
conditions. Fig. 5 reveals this phenomenon clearly.
\begin{figure}[tbh]
\epsfxsize=4in
\par
\epsfclipon
\par
\centerline{\epsffile{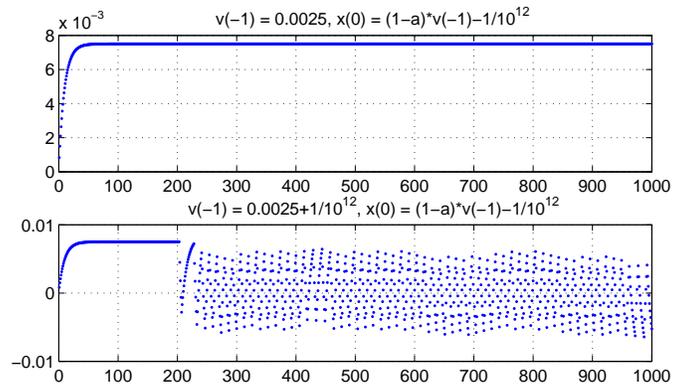}}
\caption{Sensitive dependence on initial conditions}
\end{figure}
Is the trajectory in the lower part of Fig. 5 aperiodic? Fig. 6 is its
spectrum produced using the function ``pmtm'' in Matlab. One can see that
this trajectory contains a broad band of frequencies.
\begin{figure}[tbh]
\epsfxsize=4in
\par
\epsfclipon
\par
\centerline{\epsffile{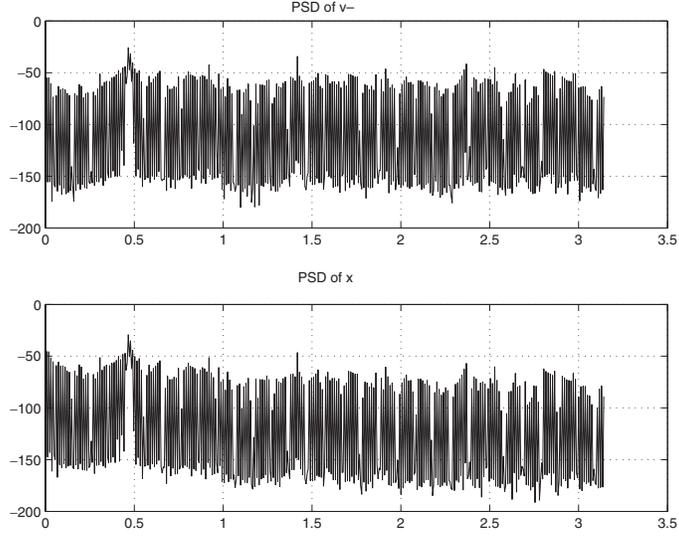}}
\caption{Spectrum of an aperiodic orbit}
\end{figure}

Next let $a=1$ and $b=-3/10$, and we get Figs. 7-8 where the
horizontal axis denotes $v(k-1)$ and the vertical axis stands for
$x(k)$. The first two (in Fig. 7) are eventually periodic orbits
of different periods, the third one (in Fig. 8) is aperiodic.
\begin{figure}[tbh]
\epsfxsize=4in
\par
\epsfclipon
\par
\centerline{\epsffile{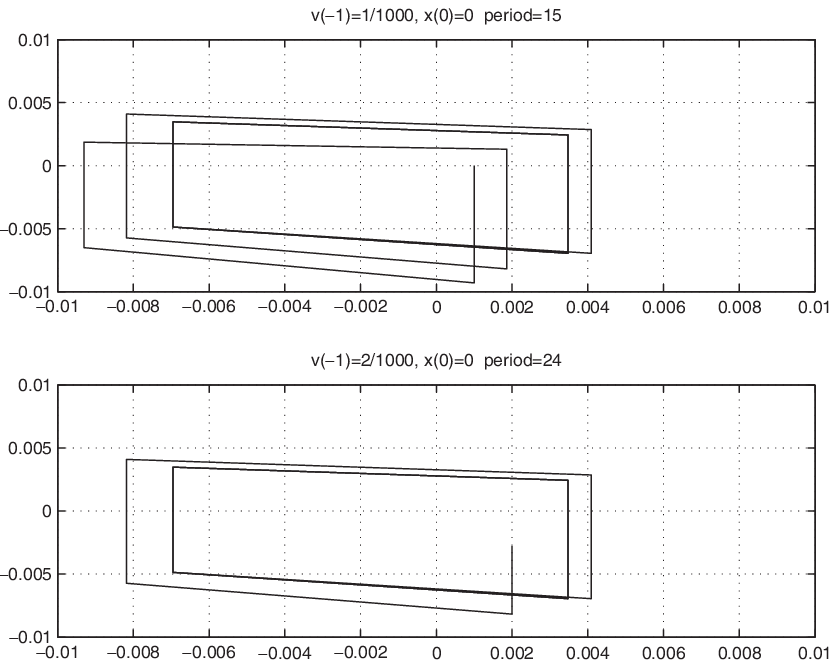}}
\caption{Two periodic orbits}
\end{figure}
\begin{figure}[tbh]
\epsfxsize=4in
\par
\epsfclipon
\par
\centerline{\epsffile{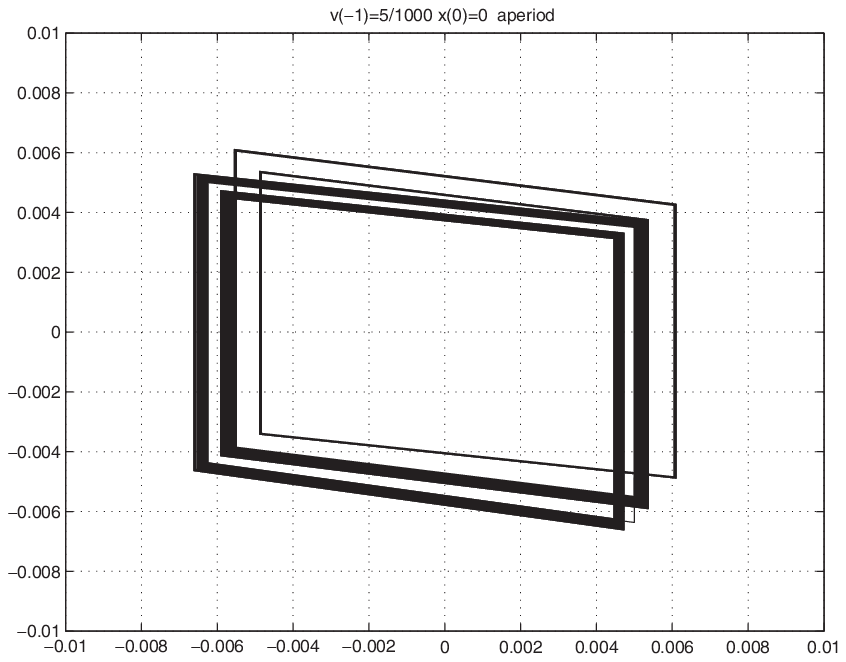}}
\caption{An aperiodic orbit}
\end{figure}

The complicated behavior of the system in Fig. 2 is due to its
nonlinearity. To some extent, invariant sets provide some measure
of how complex the dynamics of a system is. According to the above
examples, the invariant sets of the system in Eq. (\ref{exap1})
contain not only the origin, non-origin fixed points (Fig. 3), but
also periodic (Fig. 7) and aperiodic orbits (Fig. 8). Furthermore,
it may contain a strange attractor if chaos is indeed present in
the system. In the rest of this subsection, we will analyze the
dynamics of this system. We always assume that $\left| a+b \right|
<1$ which guarantees the boundedness of trajectories of the
system.

\subsubsection{Case 1:\ $\left| a\right| <1$}

For convenience, define
\begin{equation*}
\xi \left( k\right) :=\left[
\begin{array}{c}
v\left( k-1\right) \\
x\left( k\right)%
\end{array}%
\right] ,
\end{equation*}%
then the system can be written as
\begin{eqnarray}
\xi \left( k+1\right) &=&\left[
\begin{array}{cc}
1 & 0 \\
b & a%
\end{array}%
\right] \xi \left( k\right) +s_{k}\left[
\begin{array}{cc}
-1 & 1 \\
-b & b%
\end{array}%
\right] \xi \left( k\right)  \notag \\
&:=&\left( A+s_{k}B\right) \xi \left( k\right) :=F\left( \xi
\left( k\right) \right) ,\ \forall k\geq 0,  \label{case1-system}
\end{eqnarray}%
and
\begin{eqnarray}
s_{k} &=&1\mbox{ if }\left| x\left( k\right) -v\left( k-1\right) \right|
>\delta ;  \notag \\
s_{k} &=&0\mbox{ if }\left| x\left( k\right) -v\left( k-1\right) \right|
\leq \delta .  \label{case1-switch}
\end{eqnarray}

Based on this representation, the fixed points of the system are the line
segment:
\begin{equation}
x=\frac{b}{1-a}v_{-},  \label{case1-fixed-points}
\end{equation}%
within the region:
\begin{equation}
\left| x-v_{-}\right| \leq \delta .  \label{case1-bound-for-firxe-point}
\end{equation}%
(Note $v_{-}$ indicates that $v$ is one step behind $x$.) For the
local stability of fixed points, we have the following result.

\begin{proposition}
For the system in Eq. (\ref{case1-system}) with \ $\left| a\right|
<1$, a local stability region, denoted by $R_{loc}\subset
\mathbb{R}^{2}$, of its fixed
points is the region encircled by%
\begin{equation}
\left| x-v_{\_}\right| =\delta ,  \label{case1-R1-a}
\end{equation}%
and%
\begin{equation}
\left| v_{\_}\right| =\frac{1-\left| a\right| }{1-\left( a+b\right) }\delta .
\label{case1-R1-b}
\end{equation}
\end{proposition}

\noindent \textbf{Proof: } Given an initial point $\left( v\left( -1\right)
,x\left( 0\right) \right) \in R_{loc}$, we have%
\begin{equation*}
x\left( 1\right) =ax\left( 0\right) +bv\left( -1\right) .
\end{equation*}%
In general,
\begin{equation}
x\left( k\right) =a^{k}x\left( 0\right) +\sum_{i=0}^{k-1}a^{i}bv\left(
-1\right) ,  \label{case1-general-form}
\end{equation}%
provided that%
\begin{equation}
\left| x\left( k\right) -v\left( -1\right) \right| \leq \delta ,\
\forall k>0.  \label{case1-constraint}
\end{equation}%
Now we show that Eq. (\ref{case1-constraint}) indeed holds.

Since%
\begin{eqnarray*}
x\left( k\right) -v\left( -1\right) &=&a^{k}x\left( 0\right)
+\sum_{i=0}^{k-1}a^{i}bv\left( -1\right) -v\left( -1\right) \\
&=&a^{k}\left( x\left( 0\right) -v\left( -1\right) \right) +\left(
1-a^{k}\right) \frac{b+a-1}{1-a}v\left( -1\right) ,
\end{eqnarray*}%
one has%
\begin{equation*}
\left| x\left( k\right) -v\left( -1\right) \right| \leq \left| a^{k}\right|
\left| x\left( 0\right) -v\left( -1\right) \right| +\left( 1-a^{k}\right)
\frac{1-\left( a+b\right) }{1-a}\left| v\left( -1\right) \right| .
\end{equation*}%
If $0\leq a<1$, then
\begin{eqnarray*}
\left| x\left( k\right) -v\left( -1\right) \right| &\leq &a^{k}\delta
+\left( 1-a^{k}\right) \frac{1-\left( a+b\right) }{1-a}\frac{1-a}{1-\left(
a+b\right) }\delta \\
&=&\delta .
\end{eqnarray*}%
If $-1<a<0$ and $a^{k}>0$, then%
\begin{eqnarray*}
\left| x\left( k\right) -v\left( -1\right) \right| &\leq &a^{k}\delta
+\left( 1-a^{k}\right) \frac{1-\left( a+b\right) }{1-a}\frac{1+a}{1-\left(
a+b\right) }\delta \\
&=&\left( a^{k}+\left( 1-a^{k}\right) \frac{1+a}{1-a}\right) \delta \\
&\leq &\left( a^{k}+\left( 1-a^{k}\right) \right) \delta =\delta .
\end{eqnarray*}%
If $-1<a<0$ and $a^{k}<0$, then%
\begin{eqnarray*}
\left| x\left( k\right) -v\left( -1\right) \right| &\leq &-a^{k}\delta
+\left( 1-a^{k}\right) \frac{1-\left( a+b\right) }{1-a}\frac{1+a}{1-\left(
a+b\right) }\delta \\
&=&\left( -a^{k}+\left( 1-a^{k}\right) \frac{1+a}{1-a}\right) \delta \\
&=&\frac{1+a-2a^{k}}{1-a}\delta .
\end{eqnarray*}%
Therefore it suffices to show that%
\begin{equation*}
\frac{1+a-2a^{k}}{1-a}\leq 1.
\end{equation*}%
However, it is equivalent to
\begin{equation*}
a\leq a^{k},
\end{equation*}%
which holds for $-1<a<0$ and $a^{k}<0$. By taking limit in Eq. (\ref%
{case1-general-form}) with respect to $k$, $\left( v\left( k-1\right)
,x\left( k\right) \right) $ converges to a fixed point defined by Eqs. (\ref%
{case1-fixed-points})-(\ref{case1-bound-for-firxe-point}). The
proof is completed. $\hfill $ $\blacksquare $

\bigskip Having identified a local stability region, next we will study the
following problem: Can the actual stability region of the fixed
points be larger than the region given in Proposition 1? We will
see that this problem is actually a difficult one in that it
heavily depends on system parameters. Before doing so, we first
concentrate on the ``one-dimensional
case'', i.e., the dynamics of $x$, to get the globally attracting region of $x$ of the system in Eq. (\ref%
{exap1}). We have the following result.

\begin{proposition}
The globally attracting region of $x$ is given by%
\begin{equation}
\left| x\right| \leq \frac{\left| b\right| }{1-\left| a+b\right| }\delta .
\label{case1-attractor-x}
\end{equation}%
Furthermore, it is positively invariant.
\end{proposition}

\noindent \textbf{Proof: } According to Eq. (\ref{exap1}),
\begin{eqnarray*}
x\left( 1\right) &=&ax\left( 0\right) +bv\left( 0\right) =\left( a+b\right)
x\left( 0\right) +b\left( v\left( 0\right) -x\left( 0\right) \right) , \\
x\left( 2\right) &=&ax\left( 1\right) +bv\left( 1\right) =\left( a+b\right)
^{2}x\left( 0\right) +\left( a+b\right) b\left( v\left( 0\right) -x\left(
0\right) \right) +b\left( v\left( 1\right) -x\left( 1\right) \right) , \\
&&\vdots \\
x\left( k\right) &=&\left( a+b\right) ^{k}x\left( 0\right)
+\sum_{i=0}^{k-1}\left( a+b\right) ^{i}b\left( v\left( k-1-i\right) -x\left(
k-1-i\right) \right) ,
\end{eqnarray*}%
hence%
\begin{equation}
\left| x\left( n\right) \right| \leq \left| a+b\right| ^{n}\left| x\left(
0\right) \right| +\frac{1-\left| a+b\right| ^{n}}{1-\left| a+b\right| }%
\left| b\right| \delta ,\ \forall n\geq 1.  \label{case1-attractor-x-2}
\end{equation}%
By taking limit on both sides, one gets Eq.
(\ref{case1-attractor-x}). Moreover, if
\begin{equation*}
\left| x\left( 0\right) \right| \leq \frac{\left| b\right| }{1-\left|
a+b\right| }\delta ,
\end{equation*}%
then
\begin{equation*}
\left| x\left( n\right) \right| \leq \frac{\left| b\right| }{1-\left|
a+b\right| }\delta ,\ \ \ \forall n\geq 1,
\end{equation*}%
which means that the region given by Eq. (\ref{case1-attractor-x})
is positively invariant. $\hfill $ $\blacksquare $

\bigskip Based on this observation, we are ready to derive a positive
invariant set for the system in Eq. (\ref{case1-system}).

\begin{theorem}
For the system in Eq. (\ref{case1-system}), if
\begin{equation*}
\frac{\left| b\right| }{1-\left| a+b\right| }>\frac{1-\left| a\right| }{%
1-\left( a+b\right) },
\end{equation*}%
then region defined by%
\begin{equation*}
\left| x\right| \leq \frac{\left| b\right| }{1-\left| a+b\right| }\delta
\end{equation*}%
and%
\begin{equation*}
\left| v_{-}\right| \leq \frac{\left| b\right| }{1-\left| a+b\right| }\delta
\end{equation*}%
is a positively invariant set. Otherwise, the region defined by
\begin{equation*}
\left| x\right| \leq \frac{1-\left| a\right| }{1-\left( a+b\right) }\delta
\end{equation*}%
and%
\begin{equation*}
\left| v_{-}\right| \leq \frac{1-\left| a\right| }{1-\left( a+b\right) }%
\delta
\end{equation*}%
is globally attracting, which indicates that the fixed points given by Eqs. (%
\ref{case1-fixed-points})-(\ref{case1-bound-for-firxe-point}) \
are the only invariant set of the system (For convenience, we call
such a system a generic system).
\end{theorem}

\noindent \textbf{Proof: } It readily follows from Proposition 1
and Proposition 2. $\hfill $ $\blacksquare $

\bigskip The following result is an immediate consequence of Theorem 1.

\begin{corollary}
If the system in Eq. (\ref{case1-system}) satisfies either of

\begin{itemize}
\item $a>0$ and $b>0;$

\item $a<0$ and $b<0,$
\end{itemize}

then it is a generic system.
\end{corollary}

\noindent \textbf{Proof: } Suppose $a>0$ and $b>0$. Then
\begin{equation*}
\frac{\left| b\right| }{1-\left| a+b\right| }=\frac{b}{1-\left( a+b\right) }%
\leq \frac{1-a}{1-\left( a+b\right) }=\frac{1-\left| a\right| }{1-\left(
a+b\right) }.
\end{equation*}%
Hence the system is generic. On the other hand, given $a<0$ and $b<0$,
\begin{equation*}
\frac{\left| b\right| }{1-\left| a+b\right| }=\frac{-b}{1+\left( a+b\right) }%
, \ \ \ \frac{1-\left| a\right| }{1-\left( a+b\right) }=\frac{1+a}{1-\left(
a+b\right) }.
\end{equation*}%
Since
\begin{equation*}
\frac{-b}{1+\left( a+b\right) }\leq \frac{1+a}{1-\left( a+b\right) }
\end{equation*}%
is equivalent to
\begin{equation*}
a^{2}\leq 1+b^{2},
\end{equation*}%
which says
\begin{equation*}
\frac{\left| b\right| }{1-\left| a+b\right| }\leq \frac{1-\left| a\right| }{%
1-\left( a+b\right) },
\end{equation*}%
i.e., the system is generic. $\hfill $ $\blacksquare $

\bigskip Theorem 1 tells us that, in order to have complex dynamics,
\begin{equation}
\frac{\left| b\right| }{1-\left| a+b\right| }>\frac{1-\left| a\right| }{%
1-\left( a+b\right) }  \label{case1-critical}
\end{equation}%
must be satisfied. However, this is \textit{not} a sufficient condition. For
the case when
\begin{equation*}
a=9/10,\ b=-3/10,
\end{equation*}%
(which satisfies Eq. (\ref{case1-critical})), we have already
known that the system exhibits complicated dynamics (see Figs.
3-6). However, for the case when
\begin{equation*}
a=3/10,\ b=-9/10,
\end{equation*}%
which also satisfies Eq. (\ref{case1-critical}), there is no
complex dynamic behavior, i.e., the system is generic. The
following argument provides a sufficient proof for this specific
system.

Given $\left( v\left( -1\right) ,x\left( 0\right) \right) $ satisfying%
\begin{equation*}
\left| x\left( 0\right) -v\left( -1\right) \right| >\delta ,
\end{equation*}%
one has
\begin{eqnarray*}
x\left( 1\right) &=&\left( a+b\right) x\left( 0\right) , \\
v\left( 0\right) &=&x\left( 0\right) .
\end{eqnarray*}%
Suppose
\begin{equation*}
\left| x\left( 1\right) -v\left( 0\right) \right| >\delta ,
\end{equation*}%
then%
\begin{equation}
\left| x\left( 0\right) \right| >\frac{\delta }{1-\left( a+b\right) },
\label{case-inequal1}
\end{equation}%
and
\begin{eqnarray*}
x\left( 2\right) &=&\left( a+b\right) x\left( 1\right) =\left( a+b\right)
^{2}x\left( 0\right) , \\
v\left( 1\right) &=&x\left( 1\right) =\left( a+b\right) x\left( 0\right) .
\end{eqnarray*}%
If
\begin{equation}
\left| x\left( 2\right) -v\left( 1\right) \right| \leq \delta ,
\label{case1-trivial}
\end{equation}%
and
\begin{equation*}
\left| v\left( 1\right) \right| \leq \frac{1-\left| a\right| }{1-\left(
a+b\right) }\delta ,
\end{equation*}%
then the trajectory will converge to some fixed point. Meanwhile,
\begin{equation}
\left| x\left( 0\right) \right| \leq \frac{1-\left| a\right| }{1-\left(
a+b\right) }\frac{1}{\left| a+b\right| }\delta .  \label{case-inequal2}
\end{equation}%
Note that Eq. (\ref{case1-trivial}) holds given Eq.
(\ref{case-inequal1}). Therefore, only
\begin{equation}
\frac{\delta }{1-\left( a+b\right) }\leq \frac{1-\left| a\right| }{1-\left(
a+b\right) }\frac{1}{\left| a+b\right| }\delta  \label{case-inequal3}
\end{equation}%
is required. Moreover, Eq. (\ref{case-inequal3}) is equivalent to
\begin{equation}
-b\leq 1.  \label{case-inequal4}
\end{equation}%
Systems with
\begin{equation}
a=3/10,b=-9/10,  \label{sys1}
\end{equation}%
and
\begin{equation}
a=8/10,b=-9/10,  \label{sys2}
\end{equation}%
both satisfy Eq. (\ref{case-inequal4}). However, for a sufficiently large time $%
k $, any trajectory $\left( v\left( k-1\right) ,x\left( k\right)
\right) $ governed by Eq. (\ref{sys1}) will satisfy
\begin{equation*}
\left| x\left( k\right) -v\left( k-1\right) \right| >\delta ,
\end{equation*}%
and%
\begin{equation*}
\left| x\left( k\right) \right| >\frac{\delta }{1-\left( a+b\right) }.
\end{equation*}%
consequently, it will converge to a fixed point. However, any trajectory $%
\left( v\left( k-1\right) ,x\left( k\right) \right) $ governed by Eq. (\ref{sys2}%
) violates these two conditions, predicting complex dynamics, see Fig. 11
below.

We have already analyzed three cases:

\begin{itemize}
\item $a>0$ and $b>0;$

\item $a<0$ and $b<0;$

\item $a>0$ and $b<0.$
\end{itemize}

What about the case when $a<0$ and $b>0$? Next we will prove that
such a system
is generic. It is easy to see that the transition matrix of the system in Eq. (%
\ref{exap1}) is some combination of $\left( a+b\right) ^{k}$ and
$\left( a^{m}+\sum_{i=0}^{m-1}a^{i}b\right) $ with scalar
multiplication as the
involved operation, where $k\geq 0$ and $m> 1$, since $\left| a+b\right| <1$%
, if
\begin{equation}
\left| a^{m}+\sum_{i=0}^{m-1}a^{i}b\right| <1  \label{case1-inqu1}
\end{equation}%
for all $m>1$, the state $x$ will tend to the origin unless it reaches a
fixed point. In this case, the system is generic. Via simple manipulation, Eq. (%
\ref{case1-inqu1}) is equivalent to
\begin{equation}
0\leq \left( 1-a^{m}\right) \frac{1-\left( a+b\right) }{1-a}\leq 2.
\label{case1-inqu2}
\end{equation}%
Given $a<0$ and $b>0$, define
\begin{equation*}
f\left( m\right) :=\left( 1-a^{m}\right) \frac{1-\left( a+b\right) }{1-a}, \
\ \ \forall m>1,
\end{equation*}%
then%
\begin{equation*}
f\left( m\right) \geq 0, \ \ \ \forall m>1,
\end{equation*}%
and%
\begin{equation*}
f\left( 3\right) =\max_{m>1}f\left( m\right) .
\end{equation*}%
However,
\begin{eqnarray*}
f\left( 3\right) -2 &=&\left( 1-a^{3}\right) \frac{1-\left( a+b\right) }{1-a}%
-2 \\
&=&\left( 1+a+a^{2}\right) \left( 1-\left( a+b\right) \right) -2 \\
&\leq &0.
\end{eqnarray*}%
hence, Eq. (\ref{case1-inqu2}) (then Eq. (\ref{case1-inqu1}))
holds for all $m>1$, which means the system is generic.

In the rest of this part, we will concentrate on a specific system
and study its complexity. Consider the system in Fig. 2, where
\begin{figure}[tbh]
\epsfxsize=4in
\par
\epsfclipon
\par
\centerline{\epsffile{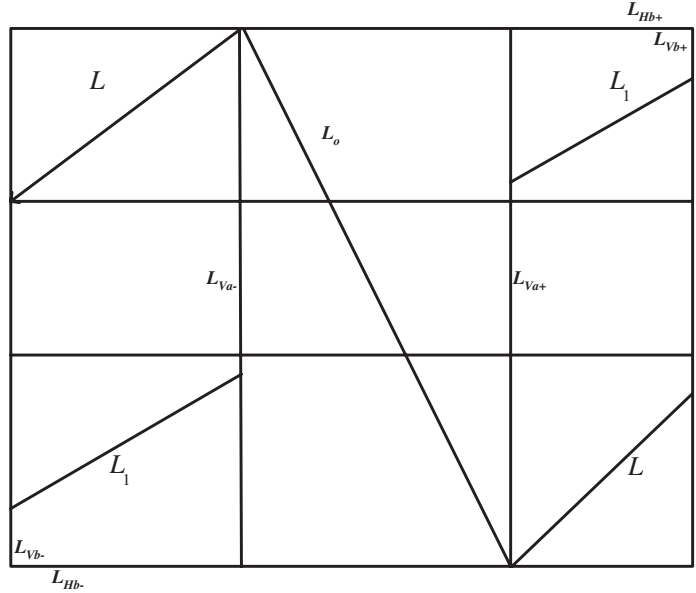}} 
\caption{Diagram for the case of $a=0.9$ and $b=-0.3$}
\end{figure}
\begin{eqnarray}
x(k+1) &=&ax(k)+bv(k),  \notag \\
u_{c}(k) &=&x(k),  \label{concrete-system}
\end{eqnarray}%
with $a=0.9$, $b=-0.3$, $v(-1)\in \mathbb{R}$, and for $k\geq 0$,
\begin{equation*}
v(k)=H_{1}\left( u_{c}\left( k\right) ,v(k-1)\right) =\left\{
\begin{array}{ll}
u_{c}(k), & \mbox{if~}\left| u_{c}\left( k\right) -v\left( k-1\right)
\right| >0.01 , \\
v(k-1), & \mbox{else.}%
\end{array}%
\right.
\end{equation*}%
We make the following definitions (Fig. 9):
\begin{eqnarray*}
L_{Hb+} &:=&\left\{ \left( v_{-},\frac{-b}{1-\left| a+b\right|
}\delta \right) :\left| v_{-}\right| \leq \frac{-b}{1-\left|
a+b\right| }\delta
\right\} , \\
L_{Hb-} &:=&\left\{ \left( v_{-},\frac{b}{1-\left| a+b\right|
}\delta \right) :\left| v_{-}\right| \leq \frac{-b}{1-\left|
a+b\right| }\delta
\right\} , \\
L_{Vb+} &:=&\left\{ \left( \frac{-b}{1-\left| a+b\right| }\delta
,x\right)
:\left| x\right| \leq \frac{-b}{1-\left| a+b\right| }\delta \right\} , \\
L_{Vb-} &:=&\left\{ \left( \frac{b}{1-\left| a+b\right| }\delta
,x\right)
:\left| x\right| \leq \frac{-b}{1-\left| a+b\right| }\delta \right\} , \\
I_{b} &:=&\left\{ \left( v_{-},x\right) :\left| v_{-}\right| \leq \frac{-b}{%
1-\left| a+b\right| }\delta ,\left| x\right| \leq \frac{-b}{1-\left|
a+b\right| }\delta \right\} .
\end{eqnarray*}%
\begin{eqnarray*}
L_{Va+} &:=&\left\{ \left( \frac{1-\left| a\right| }{1-\left( a+b\right) }%
\delta ,x\right) ,:\left| x\right| \leq \frac{-b}{1-\left( a+b\right) }%
\delta \right\} , \\
L_{Va-} &:=&\left\{ \left( -\frac{1-\left| a\right| }{1-\left( a+b\right) }%
\delta ,x\right) ,:\left| x\right| \leq \frac{-b}{1-\left( a+b\right) }%
\delta \right\} , \\
I_{a} &:=&\left\{ \left( v_{-},x\right)\in I_{b} :\left|
v_{-}\right| \leq \frac{1-\left| a\right| }{1-\left( a+b\right)
}\delta ,\left| x\right| \leq \frac{-b}{1-\left| a+b\right|
}\delta \right\} .
\end{eqnarray*}%
\begin{eqnarray*}
L_{o} &:=&\left\{ \left( v_{-},x\right) \in I_{a}:x=\frac{b}{1-a}%
v_{-}\right\} , \\
L_{\delta +} &:=&\left\{ \left( v_{-},x\right) \in
I_{b}:x-v_{-}=\delta
\right\} , \\
L_{\delta -} &:=&\left\{ \left( v_{-},x\right) \in
I_{b}:x-v_{-}=-\delta
\right\} , \\
L_{1\delta +} &:=&\left\{ \left( v_{-},x\right) \in I_{b}:x=\left(
a+b\right) v_{-} ,v_{-}>\frac{1-\left| a\right| }{1-\left( a+b\right) }%
\delta \right\} , \\
L_{1\delta -} &:=&\left\{ \left( v_{-},x\right) \in I_{b}:x=\left(
a+b\right) v_{-} ,v_{-}<-\frac{1-\left| a\right| }{1-\left( a+b\right) }%
\delta \right\} .
\end{eqnarray*}

Clearly, $L_{o}$ is the set of equilibria, $I_{a}$ is a local
stability region of $L_{o}$, and $I_{b}$ is a globally attracting
region and is also
positively invariant. Denote the two endpoints of $L_{o}$ by $E^{+}$ and $%
E^{-}$, i.e., $E^{+}=\left( -\frac{1-\left| a\right| }{1-\left( a+b\right) }%
\delta ,-\frac{1-\left| a\right| }{1-\left( a+b\right) }\delta +\delta
\right) $ and $E^{-}=\left( \frac{1-\left| a\right| }{1-\left( a+b\right) }%
\delta ,\frac{1-\left| a\right| }{1-\left( a+b\right) }\delta -\delta
\right) $. Define%
\begin{equation*}
E_{s}:=L_{o}\backslash \left\{ E^{+},E^{-}\right\} :=\left\{ \left(
v_{-},x\right) \in L_{o}:\left( v_{-},x\right) \notin \left\{
E^{+},E^{-}\right\} \right\} .
\end{equation*}%
Then each point in $E_{s}$ is stable in the sense of Lyapunov, however it is
\textit{not} asymptotically. As for the stability of $E^{+}$ (resp. $E^{-}$%
), each trajectory starting from a point in $I_{b}$ on
$v_{-}=\frac{1-\left| a\right| }{1-\left( a+b\right) }\delta $
(resp. $v_{-}=-\frac{1-\left| a\right| }{1-\left( a+b\right)
}\delta $) will converge to $E^{+}$ (resp. $E^{-}$). How about
trajectories starting from points in $I_{b}\backslash I_{a}$
sufficiently
close to $E^{+}$ (resp. $E^{-}$)? It turns out that they never converge to $%
E^{+}$ (or $E^{-}$); therefore the two equilibria $E^{+}$ (resp.
$E^{-}$) are \textit{not} stable. To wit, we need more
preparations.

For convenience, we regard the system in Eq.
(\ref{concrete-system}) as a map,
i.e., adopt the notation defined in Eq. (\ref{case1-system}):%
\begin{equation*}
\xi \left( k+1\right) =F\left( \xi \left( k\right) \right) .
\end{equation*}%
Given a set $\Omega \subset I_{b}$, define
\begin{equation}
\mbox{pre}^{n}\left( \Omega \right) :=\left\{ \left( v_{-},x\right) \in
I_{b}:F^{n}\left( \left( v_{-},x\right) \right) \subset I_{b}\right\} , \ \
\forall n\geq 0,  \label{preimage}
\end{equation}%
where $F^{0}\left( \left( v_{-},x\right) \right) =\left( v_{-},x\right) $, iteratively $%
F^{n}\left( \left( v_{-},x\right) \right) =F^{n-1}\left( \left(
v_{-},x\right) \right) $ for $n\geq 1$.

Then%
\begin{eqnarray*}
L_{\delta -}\backslash \left\{ \left( \frac{1-\left| a\right| }{1-\left(
a+b\right) }\delta ,\frac{1-\left| a\right| }{1-\left( a+b\right) }\delta
-\delta \right) \right\} &\subset &\mbox{pre}^{2}\left( L_{1\delta -}\right)
, \\
L_{\delta -}\backslash \left\{ \left( -\frac{1-\left| a\right| }{1-\left(
a+b\right) }\delta ,-\frac{1-\left| a\right| }{1-\left( a+b\right) }\delta
+\delta \right) \right\} &\subset &\mbox{pre}^{2}\left( L_{1\delta -}\right)
.
\end{eqnarray*}%
Based on this observation, we have%
\begin{equation*}
F\left( I_{b}\backslash \left\{ L_{Va-}\cup L_{Va+}\right\} \right) \subset
I_{b}\backslash \left\{ L_{Va-}\cup L_{Va+}\right\} ,
\end{equation*}%
i.e., $I_{b}\backslash \left\{ L_{Va-}\cup L_{Va+}\right\} $ is positively
invariant. As a result, trajectories starting from points in $%
I_{b}\backslash \left\{ L_{Va-}\cup L_{Va+}\right\} $, no matter how close
to $E^{+}$ (resp. $E^{-}$) they are, will \textit{not} converge to $E^{+}$
(resp. $E^{-}$), indicating that neither $E^{+}$ nor $E^{-}$ is locally
stable.

Moreover, for a given set $\Omega \subset I_{b}\backslash \left\{
L_{Va-}\cup L_{Va+}\right\} $, define%
\begin{eqnarray*}
\mbox{Img}^{n}\left( \Omega \right) &:=&\left\{ F^{n}\left( \Omega
\right)
\right\} , \\
\Psi \left( \Omega \right) &:=&\cup _{n=0}^{\infty
}\mbox{Img}^{n}\left( \Omega \right) ,
\end{eqnarray*}%
then it is easy to verify that
\begin{eqnarray*}
F\left( \Psi \left( L_{1\delta +}\right) \right) &\subset &\Psi \left(
L_{1\delta +}\right) , \\
F\left( \Psi \left( L_{1\delta -}\right) \right) &\subset &\Psi \left(
L_{1\delta -}\right) ,
\end{eqnarray*}%
which furthermore imply all trajectories starting within $I_{b}\backslash
\left\{ L_{Va-}\cup L_{Va+}\right\} $ will eventually move along the line
segment $\Psi \left( L_{1\delta +}\right) =\Psi \left( L_{1\delta -}\right) $%
. For a point $\xi \in I_{b}\backslash \left\{ L_{Va-}\cup L_{Va+}\right\} $%
, let $\omega \left( x\right) $ denote its $\omega -$limit set, define%
\begin{equation*}
\omega \left( I_{b}\backslash \left\{ L_{Va-}\cup L_{Va+}\right\} \right)
:=\left\{ \cup _{\xi \in I_{b}\backslash \left\{ L_{Va-}\cup L_{Va+}\right\}
}\omega \left( \xi \right) \right\} ,
\end{equation*}%
then
\begin{equation*}
\omega \left( I_{b}\backslash \left\{ L_{Va-}\cup L_{Va+}\right\} \right)
\subset \Psi \left( L_{1\delta +}\right) .
\end{equation*}%
Obviously%
\begin{equation*}
\omega \left( I_{a}\right) =L_{o}.
\end{equation*}%
Thus we get a characterization of the $\omega -$limit sets of the system in Eq. (%
\ref{concrete-system}). However, we have to admit that this
characterization is somewhat crude because all trajectories starting within $%
I_{b}\backslash \left\{ L_{Va-}\cup L_{Va+}\right\} $ will eventually move
along merely a\ part of each line segment in $\Psi \left( L_{1\delta
+}\right) $ instead of the whole line segment. Fig. 10 given later will
visualize this observation. Now the problem of finding the exact $\omega -$%
limit set of the system is still under our study. Nevertheless,
adopting the argument on pp. 24 in Robinsion [1995], it is easy,
though not straightforward due to the nature of the map $F$, to
show that this $\omega - $limit set is indeed \textit{invariant}.
By extensive simulation, we find that this $\omega -$limit set is
also \textit{topological transitive}, however, up to now we have
not been able to build solid theoretic background to support it.

Based on the above analysis, it is fair to say that the dynamics
of the system in Eq. (\ref{concrete-system}) is remarkably
complicated: It indeed exhibits the feature of sensitive
dependence on initial conditions, this
sensitivity locates only on $\cup _{n=0}^{\infty }\left( \mbox{pre}%
^{n}\left( L_{\delta -}\right) \cup \mbox{pre}^{n}\left( L_{\delta +}\right)
\right) $, a subset of $\cup _{n=0}^{\infty }\left( \mbox{pre}^{n}\left(
L_{1\delta -}\cup L_{1\delta +}\right) \right) $. Hence it is weakly
chaotic. Next we will calculate its generalized topological entropy in the
spirit of Kopf [2000] and Galatolo [2003].

Denote by $I_{binv+}$ the region encircled by the lines $L_{Va+}$, $L_{Vb+}$%
, $L_{1\delta +}$, $\text{Img}^{1}(L_{\delta -})$. Similarly
denote the region
encircled by the lines $L_{Va-}$, $L_{Vb-}$, $L_{1\delta -}$, $%
\text{Img}^{1}(L_{\delta +})$ by $I_{binv-}$, based on the above
analysis, we have the following claim:

\noindent \textbf{Claim 1: }The steady state of the system will settle in
the region $I_{binv+}\cup $ $I_{binv-}$.

This claim is a straightforward application of the foregoing
analysis, however it plays an important role in the calculation of
the topological entropy of the system.

For the definition of topological entropy for piecewise monotone
transformations with discontinuities, please refer to Kopf [2000].
Now we will give a construction in order to compute the
topological entropy for our system, which is clearly piecewise
monotone (under some metric defined on the system rather than
under the usual Euclidean metric; however, this is not essential.)
with discontinuities.

Define%
\begin{eqnarray}
\mbox{Pim}_{F}\left( 0\right) &:=&\left\{ L_{\delta -},L_{\delta
+}\right\} ,
\notag \\
\mbox{Pim}_{F}\left( 1\right) &:=&\left\{ L:L\cap
\mbox{Pim}_{F}\left( 0\right) =\phi ,F\left( L\right) \subset
\mbox{Pim}_{F}\left( 0\right)
\right\} ,  \notag \\
&&\vdots  \notag \\
\mbox{Pim}_{F}\left( m\right) &:=&\left\{ L:L\cap \left( \cup _{i=0}^{m-1}%
\mbox{Pim}_{F}\left( 0\right) \right) =\phi ,F^{m}\left( L\right) \subset %
\mbox{Pim}_{F}\left( 0\right) \right\} , \ \ \ m\geq 1,
\label{case1-discont}
\end{eqnarray}%
where $\phi $ stands for the empty set. Note that the elements in each $%
\mbox{Pim}_{F}\left( m\right) $ are line segments.

Denote by $\#\left( \mbox{Pim}_{F}\left( m\right) \right) $ the number of
elements in $\mbox{Pim}_{F}\left( m\right) $.

\ Before calculating the topological entropy, we need to pay a bit more
attention to the mapping $F$. Clearly, according to Fig. 9 there exists a
positive integer $M$ such that
\begin{eqnarray}
\mbox{pre}^{M+l}\left( L_{\delta +}\right) \cap L_{1\delta -} &\neq &\phi ,
\notag \\
\mbox{pre}^{M+l}\left( L_{\delta -}\right) \cap L_{1\delta +} &\neq &\phi ,\
\ \forall l\geq 1.  \label{intersection}
\end{eqnarray}

In fact, each set of intersections contains exactly one element
(one line segment). For each
given integer $n>0$, define%
\begin{equation*}
\#F(n):=\sum_{m=0}^{n}\#\left( \mbox{Pim}_{F}\left( m\right) \right) ,
\end{equation*}%
and define the \textit{topological entropy} of $F$ as%
\begin{equation}
\aleph \left( F\right) :=\lim_{n\rightarrow \infty }\frac{\log \#F(n)}{n},
\label{entropy}
\end{equation}
which is well-defined (see the proof below). Then we have

\begin{theorem}
For the system in Eq. (\ref{concrete-system}), the following
statements hold:

\begin{itemize}
\item For $m\leq M$,
\begin{equation}
\#\left( \rm{Pim}_{F}\left( m\right) \right) =2. \label{cae1-en1}
\end{equation}

\item For $m>M$,
\begin{equation}
\#\left( \rm{Pim}_{F}\left( m\right) \right) =2+2\cdot \left(
m-M\right) , \label{case1-en2}
\end{equation}%
and
\begin{equation}
\aleph \left( F\right) =0.  \label{case1-en4}
\end{equation}
\end{itemize}
\end{theorem}

\noindent \textbf{Proof:} Eq. (\ref{cae1-en1}) is self-evident, Eq. (\ref{case1-en2}%
) follows from Claim 1 restricting $\#\left( \mbox{Pim}_{F}\left( m\right)
\right) $ on $I_{binv+}\cup $ $I_{binv-}$ for $m>M$ and the analysis above. Then for sufficiently large $n$ (%
$n\geq M$),
\begin{equation*}
\#F(n)=2\left( M+1\right) +2\frac{\left( n-M\right) \left( n-M+1\right) }{2},
\end{equation*}%
thus
\begin{eqnarray*}
\aleph \left( F\right) &:=&\lim_{n\rightarrow \infty }\frac{\log
\#F(n)}{n}
\\
&=&\lim_{n\rightarrow \infty }\frac{\log \left( 2\left( M+1\right) +\left(
n-M\right) \left( n-M+1\right) \right) }{n} \\
&=&0.
\end{eqnarray*}
$\hfill$ $\blacksquare$

\bigskip \noindent \textbf{Remark 2: } In light of this result, from the
perspective of topological entropy, our system is a weakly chaotic system.

The above discussion is mainly for the case of $\left| b\right|
<a$. For example, given $a=0.9$ and $b=-0.3$, Fig. 10 plots a
trajectory at large time instants, i.e., its asymptotic behavior.
Now consider the case when $a=0.8$ and $b=-0.9$, hence $\left|
b\right| <a$, and we also draw its asymptotic behavior in Fig. 11
from the same initial point. \ We observe that their asymptotic
behavior is different. The reason is still unclear up to now.
\begin{figure}[tbh]
\epsfxsize=4in
\par
\epsfclipon
\par
\centerline{\epsffile{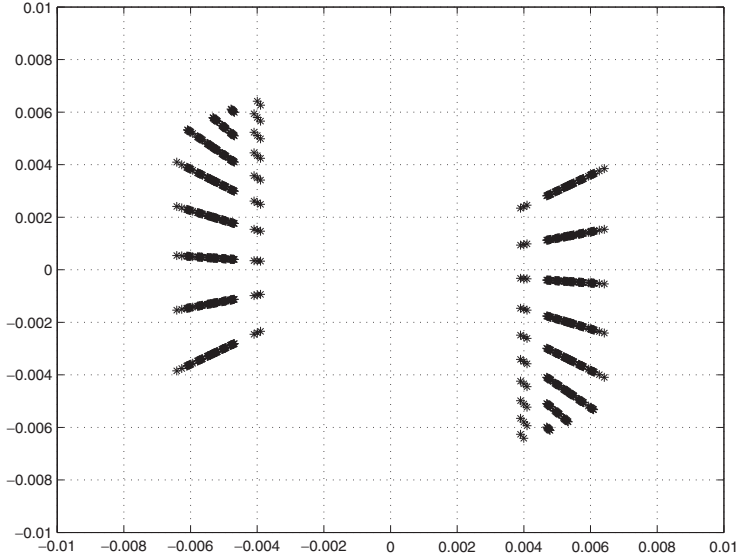}}
\caption{A trajectory at large time instants for $\left| b\right| <a$}
\end{figure}
\begin{figure}[tbh]
\epsfxsize=4in
\par
\epsfclipon
\par
\centerline{\epsffile{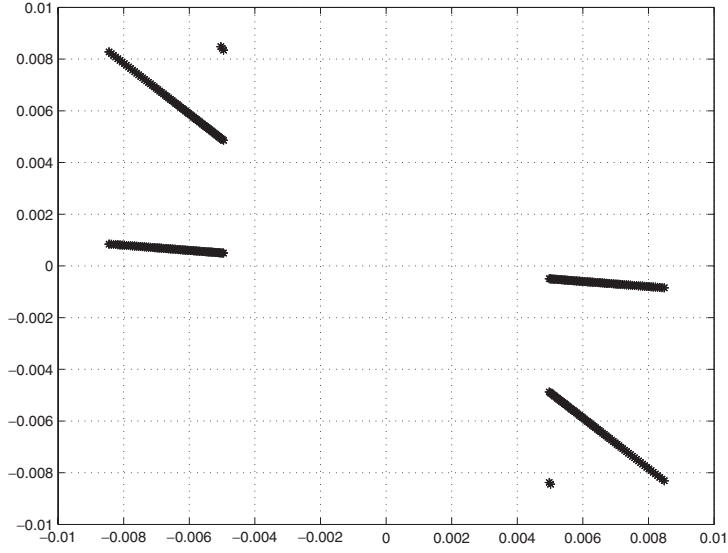}}
\caption{A trajectory at large time instants $\left| b\right| >a$}
\end{figure}

\subsubsection{\protect\bigskip Case 2:\ $a=1$}

Consider the system
\begin{eqnarray}
x(k+1) &=&x(k)+bv(k),  \notag \\
u_{c}(k) &=&x(k),  \label{case2-concrete-system}
\end{eqnarray}%
where $\left| 1+b\right| <1$. Let $v(-1)\in \mathbb{R}$, and for
$k\geq 0$,
\begin{equation*}
v(k)=H_{1}\left( u_{c}\left( k\right) ,v(k-1)\right) =\left\{
\begin{array}{ll}
u_{c}(k), & \mbox{if~}\left| u_{c}\left( k\right) -v\left( k-1\right)
\right| >0.01, \\
v(k-1), & \mbox{otherwise.}%
\end{array}%
\right.
\end{equation*}%
Figs. 7-8 show that the dynamics of the system in Eq. (\ref%
{case2-concrete-system}) can be fairly complicated. In the rest of
this subsection we mainly study the problem when the system will
have periodic orbits. From now on, we assume
\begin{equation*}
-1\leq b<0.
\end{equation*}%
For this case, $L_{Va-}$, $L_{Va+}$ and $L_{o}$ in Fig. 9 now become one
line segment
\begin{equation*}
L:=\left\{ \left( 0,x\right) :\left| x\right| \leq \delta \right\} .
\end{equation*}%
Define
\begin{equation*}
\Gamma _{in}:=\left\{ \left( v_{-},x\right) \in I_{b}:\left| x-v_{-}\right|
\leq \delta \right\} ,
\end{equation*}%
and%
\begin{equation*}
\Gamma _{ex}:=I_{b}\backslash \Gamma _{in},
\end{equation*}%
the following is a necessary condition for the existence of
periodic orbits.

\begin{theorem}
If the system in Eq. (\ref{case2-concrete-system}) has periodic
orbits, then there exist an even integer $n>0$ and integers
$K_{i}>0$ such that
\begin{equation}
\prod_{i=1}^{n}\left( 1+K_{i}b\right) =1.  \label{case2-necessary}
\end{equation}
\end{theorem}

Without loss of generality, we here prove the case of $n=2$. The
following Lemma is used in the proof of Theorem 3:

\begin{lemma}
Suppose $\xi _{0}\in \Gamma _{ex}$ is a point on a periodic orbit at time $%
K_{0}$, it will be inside $\Gamma _{in}$ at $K_{0}+1$.
\end{lemma}

\noindent \textbf{Proof: } Let $\xi _{0}=\left[
\begin{array}{c}
v_{-} \\
x%
\end{array}%
\right] $. Then
\begin{equation*}
\xi _{1}=\left[
\begin{array}{c}
v_{1} \\
x_{1}%
\end{array}%
\right] ,
\end{equation*}%
where,
\begin{eqnarray*}
x_{1} &=&\left( a+b\right) x, \\
v_{1} &=&x.
\end{eqnarray*}%
Hence
\begin{equation*}
\left| x_{1}-v_{1}\right| =\left| 1-\left( a+b\right) \right|
\left| x\right| =\left| b\right| \left| x\right|.
\end{equation*}%
Since $\left[
\begin{array}{cc}
v_{-} & x%
\end{array}%
\right] ^{^{\prime }}$ is the steady state (a periodic point),
\begin{equation*}
\left| x\right| \leq \frac{\left| b\right| \delta }{1-(a+b)}=\delta .
\end{equation*}%
One obtains
\begin{equation*}
\left| x_{1}-v_{1}\right| =\left| 1-\left( a+b\right) \right| \left|
x\right| \leq \left| b\right| \delta \leq \delta ,
\end{equation*}%
i.e., $\xi _{1}\in \Gamma _{in}$. $\hfill $ $\blacksquare $

\bigskip Note that Lemma 2 is not trivial because there are systems such as the
case when $a=3/10$ and $b=-9/10$ violating this property.

\noindent \textbf{Proof of Theorem 3: } Without loss of generality, suppose
the periodic orbit begins with $\xi _{0}\in \Gamma _{ex}$ , and by Lemma 2,
\begin{equation}
\xi _{1}=A_{1}\xi _{0}\in \Gamma _{in},  \label{first-step}
\end{equation}%
where
\begin{equation*}
A_{1}=\left[
\begin{array}{cc}
0 & 1 \\
0 & 1+b%
\end{array}%
\right] .
\end{equation*}%
Assume after a time $K_{1}$ the state
\begin{equation*}
\xi _{2}=\left( A_{1}+A_{2}\right) ^{K_{1}-1}\xi _{1}
\end{equation*}%
is in $\Gamma _{ex}$, where
\begin{equation*}
A_{2}=\left[
\begin{array}{cc}
1 & -1 \\
b & -b%
\end{array}%
\right] .
\end{equation*}%
Then
\begin{equation*}
\xi _{3}=A_{1}\xi _{2}\in \Gamma _{in}.
\end{equation*}%
After a time $K_{2}$ the state
\begin{equation*}
\xi _{4}=\left( A_{1}+A_{2}\right) ^{K_{2}-1}\xi _{3}
\end{equation*}%
returns to $\xi _{0}$. Then
\begin{equation}
\left( A_{1}+A_{2}\right) ^{K_{2}-1}A_{1}\left( A_{1}+A_{2}\right)
^{K_{1}-1}A_{1}\xi _{0}=\xi _{0}.  \label{periodic}
\end{equation}%
By straightforward algebraic computations, one gets
\begin{eqnarray}
dcx &=&x,  \label{period-equation} \\
dx &=&v_{-},  \notag
\end{eqnarray}%
where
\begin{equation*}
d=1+K_{2}b,\ c=1+K_{1}b.
\end{equation*}%
Since $\xi _{0}$ is outside the sector,
\begin{equation*}
\left| x-v_{-}\right| >\delta .
\end{equation*}%
According to Eq. (\ref{period-equation}), note that $c\neq 0$,
$d\neq 0$, then
\begin{equation}
cd=1.  \label{cd}
\end{equation}%
Given $a=1$,
\begin{equation}
K_{1}+K_{2}=K_{1}K_{2}\left( -b\right) ,  \label{haha}
\end{equation}%
which is equivalent to Eq. (\ref{case2-necessary}) for $n=2$. $\hfill $ $%
\blacksquare $

\bigskip \noindent \textbf{Remark 3: } Theorem 3 provides a necessary
condition for having periodic orbits. Interestingly\ for the case of $n=2$,
extensive experiments imply that there are periodic orbits of period $%
K_{1}+K_{2}$ if $K_{1}$ and $K_{2}$ satisfy Eq.
(\ref{case2-necessary}), and there are no periodic orbits if there
are no such $K_{1}$ and $K_{2}$ that satisfy Eq.
(\ref{case2-necessary}). Based on the observation, Theorem 3 is
not severely conservative.

Following the above analysis, we immediately have

\begin{corollary}
Suppose $a$ is rational and $b$ is irrational in Eq.
(\ref{case1-system}), then there are no periodic orbits.
\end{corollary}

\noindent \textbf{Proof: } Following the proof above, it suffices to show
that
\begin{equation}  \label{ha}
\prod_{i=1}^{n}\left( a^{K_{i}}+\sum_{j=0}^{K_{i}-1}a^{j}b\right) =1
\end{equation}%
has no positive integer solutions for any given even number $n>0$.
This can be easily verified. $\hfill $ $\blacksquare $

\bigskip Though the above result is simple, its significance can not be
underestimated: If a system has periodic orbits, then it is
\textit{not} structurally stable.

Now suppose a system has periodic orbits, how to find them? And how to
determine their periods? we first consider an example.

\noindent \textbf{Example 1: } In the case when $a=1$ and $b=-0.3$, there
are two periodic solutions. One is of period $24$ corresponding to $K_{1}=4$
and $K_{2}=20$, the other is of period $15$ corresponding to $K_{1}=5$ and $%
K_{2}=10.$ Observe that%
\begin{equation*}
-b=0.3=\frac{3}{2\cdot 5}:=\frac{p}{q_{1}\cdot q_{2}},
\end{equation*}%
where $p=3$, $q_{1}=2$, $q_{2}=5$. Interestingly%
\begin{equation*}
4=\frac{q_{2}+1}{p}q_{1},\ 20=\frac{q_{2}+1}{p}q_{1\cdot }q_{2},
\end{equation*}%
\begin{equation*}
5=\frac{q_{1}+1}{p}q_{2},\ 10=\frac{q_{1}+1}{p}q_{1\cdot }q_{2}.
\end{equation*}%
Based on this observation, we propose a necessary condition for
Theorem 3 for the case when $n=2$.

\begin{theorem}
Given
\begin{equation*}
a=1\quad and\quad b=-\frac{p}{q},
\end{equation*}%
suppose positive integers $p$ and $q$ satisfy $1<p<q$, $p\neq 2$(which is
the trivial case), and $\gcd (p,q)=1$, i.e., the greatest common divisor of $%
p$ and $q$ is $1$. Define%
\begin{equation}
\Delta :=\left\{ q_{i}:q_{i}\ is\ a\ prime\ number,\ q_{i}|q \right\} .
\label{prime-number-set}
\end{equation}%
Then if $p|\left( q_{i}+1\right) $,
\begin{equation*}
\left( \frac{q_{i}+1}{p}\frac{q}{q_{i}},\frac{q_{i}+1}{p}q\right)
\end{equation*}%
is a solution of Eq. (\ref{case2-necessary}).
\end{theorem}

\noindent \textbf{Proof: } Obviously, given $p|\left( q_{i}+1\right) $, $%
\left( \frac{q_{i}+1}{p}\frac{q}{q_{i}},\frac{q_{i}+1}{p}q\right)
$ is a solution of Eq. (\ref{case2-necessary}). Now we show how
the set $\Delta $ is constructed in the above way. Given $q_{i}\in
\Delta $, If there are two
positive numbers $m$ and $n$ satisfying%
\begin{equation}
\frac{1}{m}+\frac{1}{n}=\frac{p}{q_{i}},  \label{auxilliary}
\end{equation}%
then $\left( m\frac{q}{q_{i}},mq\right) $ is a solution to Eq. (\ref%
{case2-necessary}). Hence we need only to pay attention to solutions to Eq. (%
\ref{auxilliary}). Suppose $\left( m,n\right) $ is a solution of Eq. (\ref%
{auxilliary}), then either $\gcd \left( m,q_{i}\right) =1$ or $\gcd \left(
n,q_{i}\right) =1$ (otherwise, $p=2$ or does not exist). For convenience, we
always assume $\gcd \left( m,q_{i}\right) =1$. According to Eq. (\ref{auxilliary}%
),
\begin{equation*}
\frac{m+n}{nm}=\frac{p}{q_{i}},
\end{equation*}%
i.e.,
\begin{equation*}
\left( m+n\right) q_{i}=pmn.
\end{equation*}%
Then $q_{i}|pmn$. Since $\gcd \left( p,q_{i}\right) =\gcd \left(
m,q_{i}\right) =1$, $q_{i}|n$. Let
\begin{equation}
n=kq_{i},  \label{au3}
\end{equation}%
which leads to%
\begin{equation*}
\frac{1}{m}+\frac{1}{kq_{i}}=\frac{p}{q_{i}}.
\end{equation*}%
Consequently,%
\begin{equation*}
m\left( pk-1\right) =kq_{i},
\end{equation*}%
hence $m|kq_{i}$. Since $\gcd \left( m,q_{i}\right) =1$, $m|k$. In light of Eq. (%
\ref{au3}), we set
\begin{equation*}
n=ml.
\end{equation*}%
Substituting it into Eq. (\ref{auxilliary}), one has%
\begin{equation*}
\frac{1}{m}+\frac{1}{ml}=\frac{p}{q_{i}},
\end{equation*}%
equivalently,%
\begin{equation*}
mpl=q_{i}\left( l+1\right) ,
\end{equation*}%
which means $q_{i}|mpl$. i.e., $q_{i}|l$. Similarly, $l|q_{i}\left(
l+1\right) $, hence $l=q_{i}$. Therefore,%
\begin{equation*}
m=\frac{q_{i}+1}{p}.
\end{equation*}%
If $p|\left( q_{i}+1\right) $, then
\begin{equation*}
\left( \frac{q_{i}+1}{p},\frac{q_{i}+1}{p}q_{i}\right)
\end{equation*}%
solves Eq. (\ref{auxilliary}), and
\begin{equation*}
\left( \frac{q_{i}+1}{p}\frac{q}{q_{i}},\frac{q_{i}+1}{p}q\right)
\end{equation*}%
is a solution of Eq. (\ref{case2-necessary}). $\hfill $
$\blacksquare $

\bigskip The above theorem provides a construction for the solutions to Eq. (\ref%
{case2-necessary}). However, this is somewhat inadequate. For example, for $%
a=1$ and $b=-\frac{3}{7}$, the set $\Delta $ is empty. There are
no positive integers satisfying Eq. (\ref{case2-necessary}) for
$n=2$ either. This is good for us. However for $a=1$ and
$b=-\frac{3}{2\cdot 5\cdot 11}$, we have the following
observations (Table 1):

\begin{center}
\begin{table}[tbh]
\caption{Some periodic orbits}\centering
\begin{tabular}{|c|c|c|c|c|}
\hline
$\left( v_{-},x_{0}\right) $ & $\left( \frac{7}{1000},0\right) $ & $\left(
\frac{2}{1000},0\right) $ & $\left( \frac{1}{1000},0\right) $ & $\left(
\frac{1}{2000},0\right) $ \\ \hline
Periods & $165$ & $147$ & $243$ & $480$ \\ \hline
$\left( v_{-},x_{0}\right) $ & $\left( \frac{1}{2000},\frac{1}{2000}\right) $
& $\left( \frac{1}{2000},\frac{1}{8000}\right) $ & $\left( \frac{1}{2000},%
\frac{1}{8500}\right) $ & $\left( \frac{1}{3000},0\right) $ \\ \hline
Periods & $264$ & $480$ & $243$ & $1083$ \\ \hline
\end{tabular}%
\end{table}
\end{center}

 Periods $165$, $264$ and $480$ can be obtained based on Theorem
4, however, others can not. Actually there are more periodic and
aperiodic orbits (Table 2):

\begin{center}
\begin{table}[tbh]
\caption{More periodic orbits and aperiodic orbits}\centering
\begin{tabular}{|c|c|c|c|c|}
\hline
$\left( v_{-},x_{0}\right) $ & $\left( \frac{\delta }{1000},\frac{\left(
a+b\right) \delta }{1000}\right) $ & $\left( \frac{\delta }{100},\frac{%
\left( a+b\right) \delta }{100}\right) $ & $\left( \frac{\delta }{10},\frac{%
\left( a+b\right) \delta }{10}\right) $ & $\left( \frac{2\delta }{10},\frac{%
2\left( a+b\right) \delta }{10}\right) $ \\ \hline
Periods & $4107$ & $264$ & $243$ & $165$ \\ \hline
$\left( v_{-},x_{0}\right) $ & $\left( \frac{4\delta }{10},\frac{4\left(
a+b\right) \delta }{10}\right) $ & $\left( \frac{9\delta }{20},\frac{9\left(
a+b\right) \delta }{20}\right) $ & $\left( \frac{5\delta }{10},\frac{5\left(
a+b\right) \delta }{10}\right) $ & $\left( \frac{6\delta }{10},\frac{6\left(
a+b\right) \delta }{10}\right) $ \\ \hline
Periods & aperiodic & aperiodic & $147$ & aperiodic \\ \hline
$\left( v_{-},x_{0}\right) $ & $\left( \frac{7\delta }{10},\frac{7\left(
a+b\right) \delta }{10}\right) $ & $\left( \frac{8\delta }{10},\frac{8\left(
a+b\right) \delta }{10}\right) $ & $\left( \frac{9\delta }{10},\frac{9\left(
a+b\right) \delta }{10}\right) $ & $\left( \frac{10\delta }{10},\frac{%
10\left( a+b\right) \delta }{10}\right) $ \\ \hline
Periods & $165$ & $165$ & $243$ & $4107$ \\ \hline
\end{tabular}%
\end{table}
\end{center}

Furthermore, we observed that there are at least 55 solutions to
Eq. (\ref{case2-necessary}) for $n=4$.

The foregoing analysis tells us:\
\begin{itemize}
\item There may exist periodic orbits of very large periods.

\item There are always aperiodic orbits.
\end{itemize}

Inspired by the proof of Corollary 2, especially by Eq.
(\ref{ha}), now we attempt to construct systems with $\left|
a\right| <1$ that have periodic orbits. First choose $n=2$,
$a=9/10$, choose $K_{1}=15$ and $K_{2}=7$. Then

\begin{equation*}
b=-\frac{9015229097816388767119}{41428905812371212328810}
\end{equation*}%
solves Eq. (\ref{ha}). Surprisingly the trajectory starting from
\begin{equation*}
\left( v_{-},x_{0}\right) =\left( \frac{-b\ast \delta }{1-\left| a+b\right| }%
-\frac{1}{10^{4}},0\right)
\end{equation*}%
will become a periodic orbit of period $22(=K_{1}+K_{2})$ after some
iterations, i.e., it is an eventually periodic orbit. It can be shown this
periodic orbit is locally stable. However, a trajectory starting outside the
stability region, say, from
\begin{equation*}
\left( v_{-},x_{0}\right) =\left( \frac{-b\ast \delta }{1-\left| a+b\right| }%
-\frac{1}{10^{4}},\frac{1}{10^{3}}\right)
\end{equation*}%
is aperiodic. \ For the case when $\left| a\right| >1$, suppose
$a=11/10$, choose $K_{1}=7$ and $K_{2}=5$. Then

\begin{equation*}
b=-\frac{2138428376721}{5792012767210}
\end{equation*}%
solves Eq. (\ref{ha}). And the trajectory starting from
\begin{equation*}
\left( v_{-},x_{0}\right) =\left( \frac{\delta }{10^{5}},\frac{\left(
a+b\right) \ast \delta }{10^{5}}\right)
\end{equation*}%
will become a periodic orbit of period $12(=K_{1}+K_{2})$ after some
iterations, i.e., it is an eventually periodic orbit. It can be shown this
periodic orbit is also locally stable and there are aperiodic orbits too.

\noindent \textbf{Remark 4: } From this construction, one finds
out that most systems with $\left| a\right| <1$ or $\left|
a\right| >1$ will be unlikely to have periodic orbits.

\subsubsection{Case 3:\ $\left| a\right| >1$}

This case is analogous to that of $\left| a\right| <1$ except that all the
fixed points are unstable. The Fig. 12 is one trajectory at sufficiently
large time instants.
\begin{figure}[tbh]
\epsfxsize=4in
\par
\epsfclipon
\par
\centerline{\epsffile{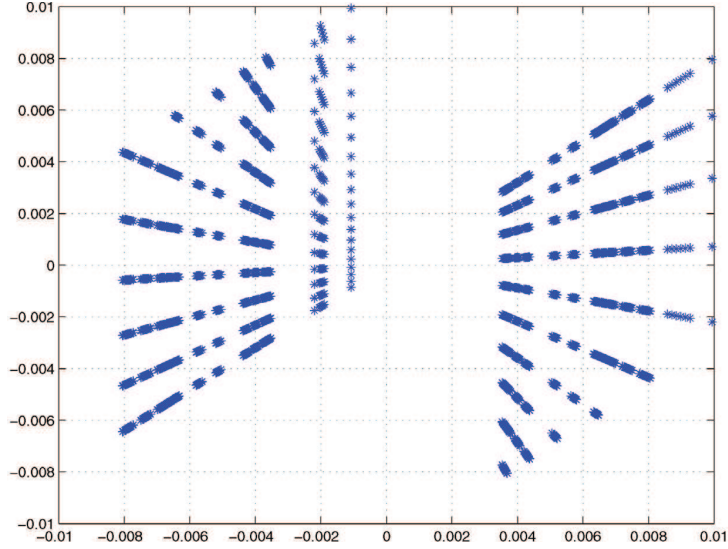}}
\caption{A trajectory at large time instants}
\end{figure}

The complex dynamics exhibited by our system is due to its
nonlinearity. This is different from a quantized system. The
complicated behavior of an unstable quantized scalar system is
extensively studied in Delchamps [1988, 1989, 1990], Fagnani \&
Zampieri [2003], etc. In Delchamps [1990], it is mentioned that\
given that the system parameter $a$ is stable, a quantized system
may have many fixed points as well as many periodic orbits which
are all asymptotically stable. However, for our constrained
systems, most of them will not possess periodic orbits. For the
systems with $a=1$, periodic orbits are locally stable, this is
not the case of for a quantized system (Delchamps [1990]). Given
that $a$ is unstable, the ergodicity of the quantized system is
studied in Delchamps [1990]. In essence, related results there
depend heavily on the affine representation of the system by which
the system is piecewise expanding, i.e., the absolute value of
derivative of the piecewise affine map in each interval is greater
than $1$. Based on this crucial property, the main theorem
(Theorem 1) in Lasota \& Yorke [1973] and then that of Li \& Yorke
[1978] are employed to show there exists a unique invariant
measure under the affine mapping and which is also ergodic with
respect to that mapping. Therefore ergodicity is established for
scalar unstable quantized systems. However, this is not the case
for our system. Though the system is piecewise linear, it is
\textbf{singular} with respect to the Lebesgue measure and
furthermore, the derivative of the system in the region $\Gamma
_{in}$ is $\left( a+b\right) $, whose absolutely value is strictly
less than $1$. Hence the results in Lasota \& Yorke [1973] and Li
\& Yorke [1978] are not applicable here. By extensive experiments,
we strongly believe that the system indeed has the property of
ergodicity, however, the problem still remains open.

To appreciate what qualitative behavior of a higher dimensional
system can have, we give the following example.

\noindent \textbf{Example 2: }Suppose the system $G$ in Fig. 2 is given by%
\begin{eqnarray*}
x\left( k+1\right) &=&2x(k)+3v(k), \\
y_{c}(k) &=&x(k),
\end{eqnarray*}%
and the controller $C$ is given by%
\begin{eqnarray*}
x_{d}\left( k+1\right) &=&-2x_{d}(k)+1.5e_{c}(k), \\
u_{c}(k) &=&x_{d}(k), \\
e_{c}(k) &=&r\left( k\right) -z\left( k\right) ,
\end{eqnarray*}%
where $v(k)$ and $z\left( k\right) $ are outputs of Eqs. (\ref{constraint1})-(%
\ref{constraint2}) respectively. We set $r\equiv 0$. We call the
resulting system $\Sigma_{o}$. It is easy to see that the
closed-loop system without the constraints $H_{1}$ and $H_{2}$ is
asymptotically stable. Under $H_{1}$ and $H_{2}$, three figures,
Figs. 13--15, are drawn. The first is for $\left( v\left(
k-1\right) ,x\left( k\right) \right) $, the second for $\left(
z\left( k-1\right) ,x_{d}(k)\right) $ and the last for $\left(
x\left( k\right) ,x_{d}(k)\right) $.
\begin{figure}[tbh]
\epsfxsize=4in
\par
\epsfclipon
\par
\centerline{\epsffile{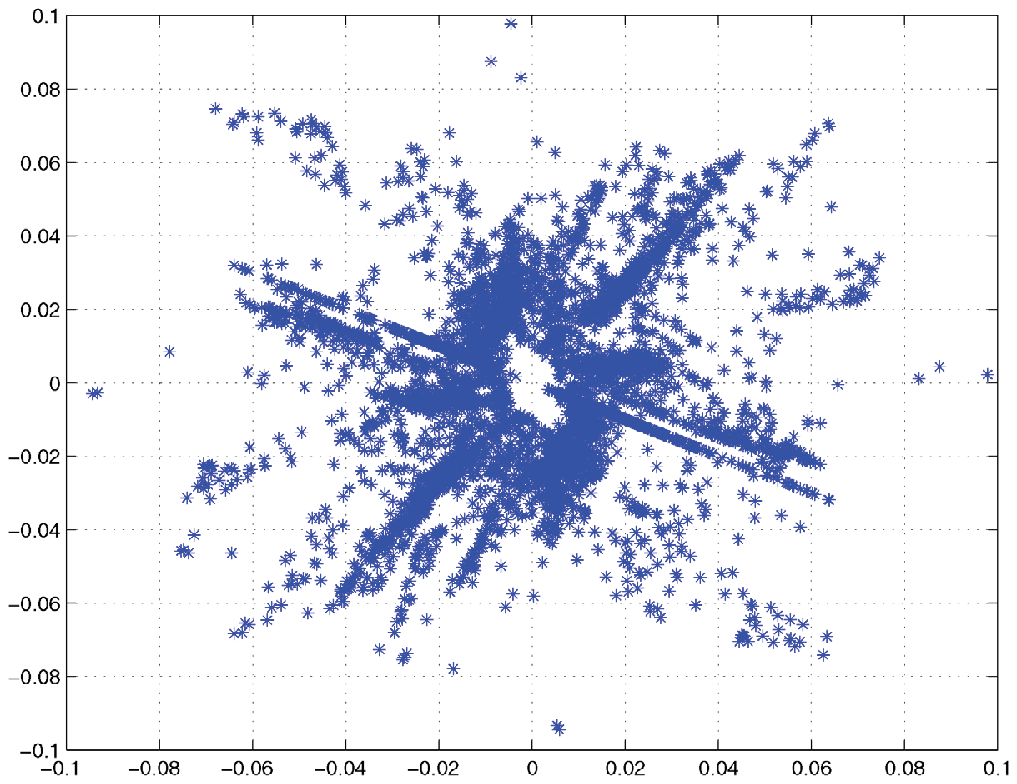}}
\caption{Plot of $\left( v\left( k-1\right) ,x\left( k\right) \right) $ at
large time instants ($\geq 95000$)}
\end{figure}
\begin{figure}[tbh]
\epsfxsize=4in
\par
\epsfclipon
\par
\centerline{\epsffile{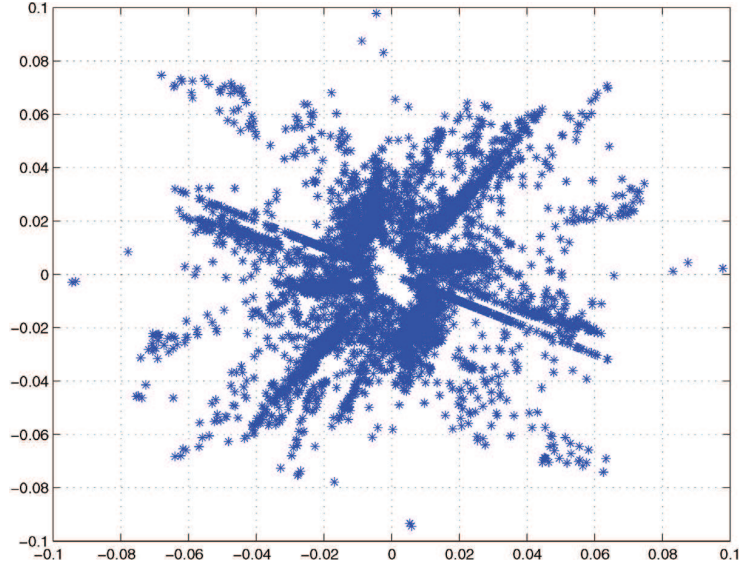}}
\caption{Plot of $\left( z\left( k-1\right) ,x_{d}(k)\right) $ at large time
instants ($\geq 95000$)}
\end{figure}
\begin{figure}[tbh]
\epsfxsize=4in
\par
\epsfclipon
\par
\centerline{\epsffile{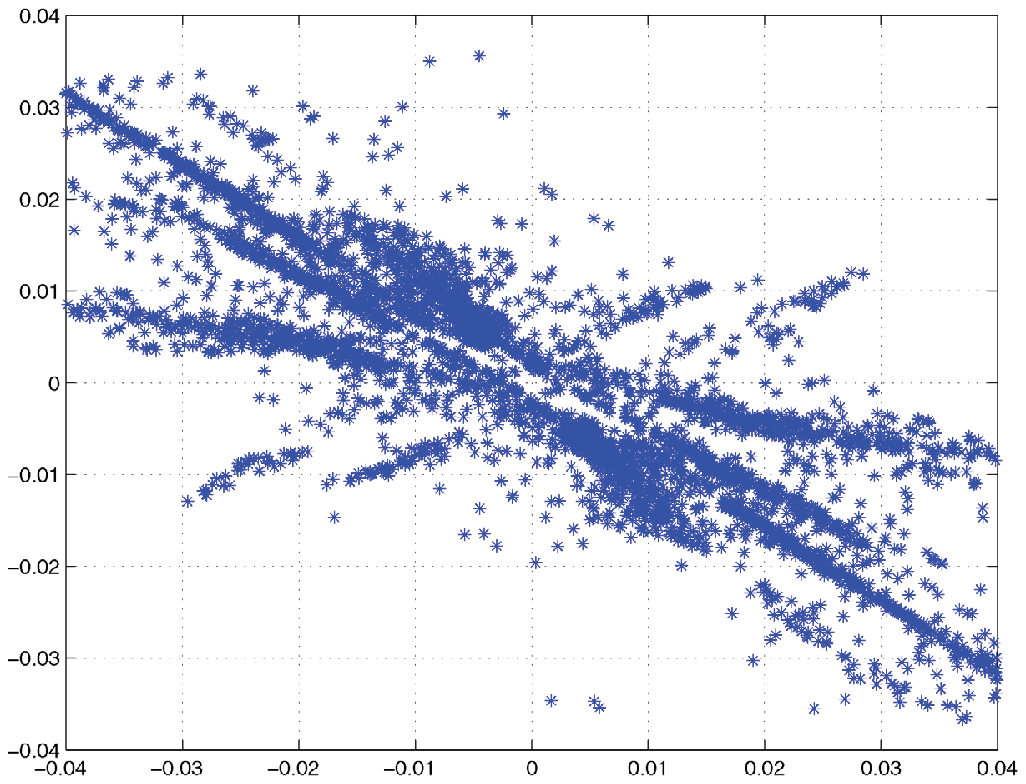}}
\caption{Plot of $\left( x\left( k\right) ,x_{d}(k)\right) $ at large time
instants ($\geq 95000$)}
\end{figure}

Next we analyze the chaotic behavior of system $\Sigma_{o}$ using
nonlinear data analysis. First we show sensitive dependence on
initial conditions. Choose an initial condition 
\[
[v(-1), x(0),
z(-1), x_{d}(0)]=[-1/1000, 1/1000, 2/1000, -1/1000],
\]
 set the
iteration number to be 600000, then we get trajectory of $x$;
perturb the initial condition above slightly to $[-1/1000,
1/1000+1/10^{13}, 2/1000, -1/1000]$, under the same iteration, we
get another trajectory of $x$, the following plot (Fig.16) is the
difference between these two $x$ of the last 1200 points of the
iteration:
\begin{figure}[tbh]
\epsfxsize=4in
\par
\epsfclipon
\par
\centerline{\epsffile{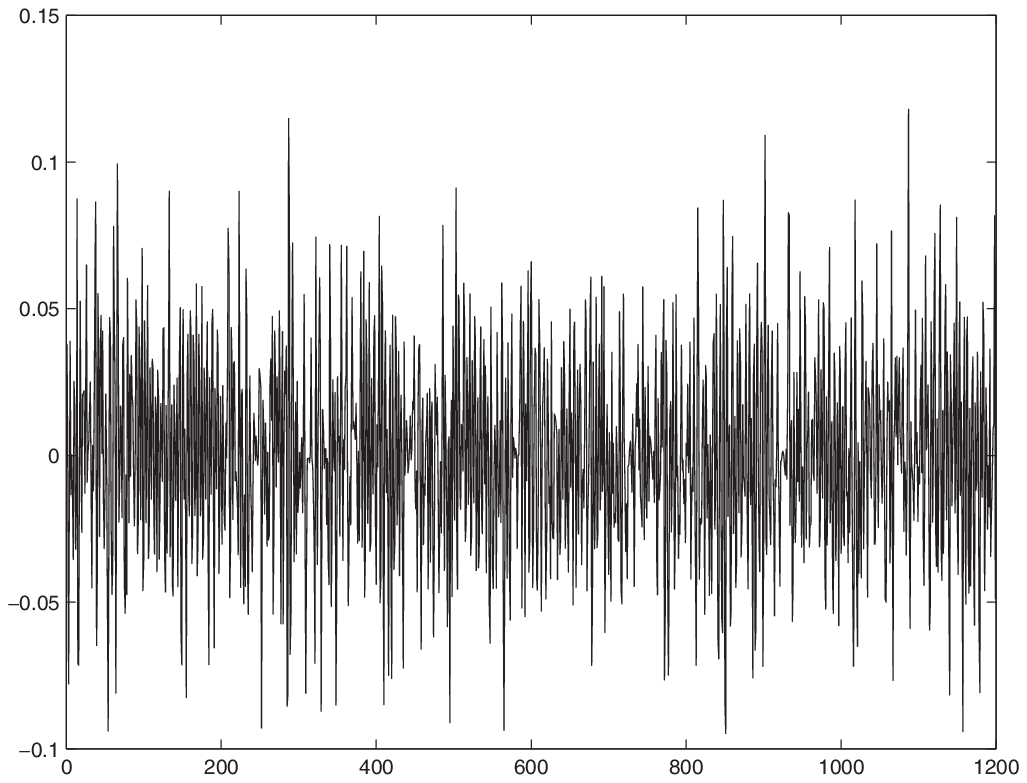}} \caption{Sensitive dependence
on initial conditions}
\end{figure}
From this figure, one can clearly see sensitive dependence on
initial conditions. In general, the spectra of a chaotic orbit
will be continuous. Here we draw the spectrum of $x$ starting from
$[-1/1000, 1/1000, 2/1000, -1/1000]$ (Fig. 17):
\begin{figure}[tbh]
\epsfxsize=4in
\par
\epsfclipon
\par
\centerline{\epsffile{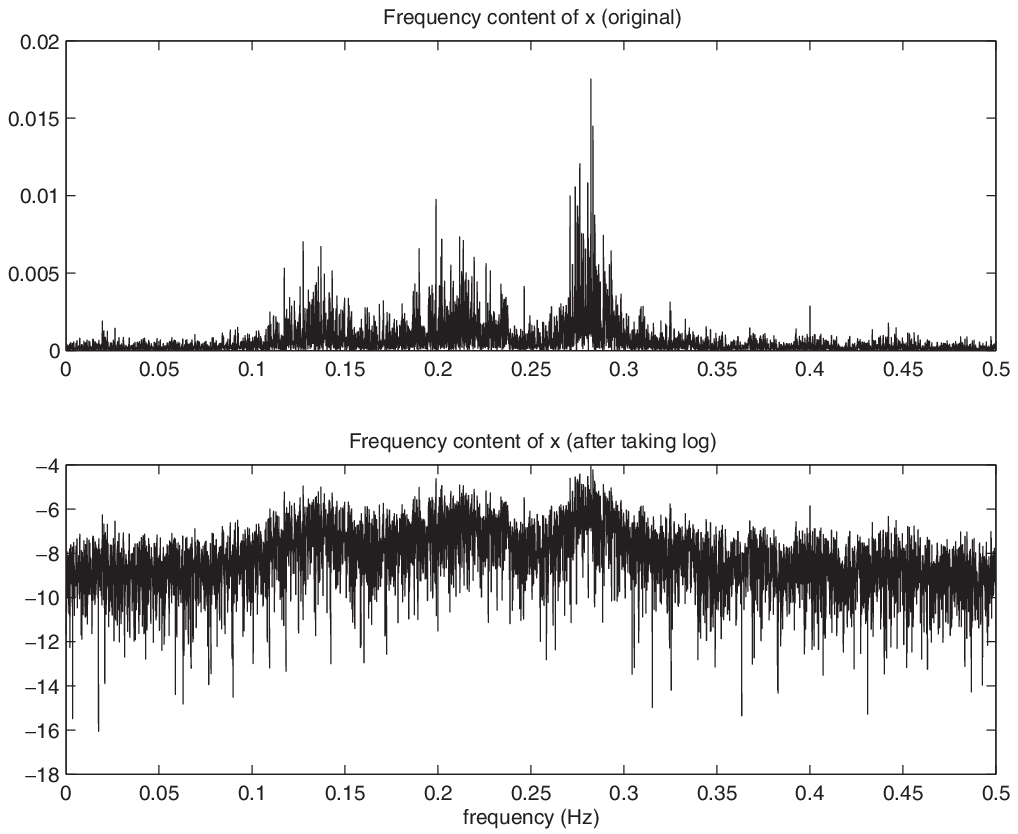}} \caption{Sensitive dependence
on initial conditions}
\end{figure}
What about the Lyapunov exponents? based on the last 10000 point
of $x$, using the software ``Chaos Data Analyzer'', choosing
parameters $D=3$, $n=3$ and $A=10^{-4}$, we get the largest
Lyapunov exponent $0.407\pm 0.027$, indicating the trajectory is
indeed a chaotic one.

Now we look at the dynamics of Example 2 geometrically. For a
given dynamical system, generally complicated manifold structure
will lead to complex dynamics. we now indicate that the manifold
structure of system $\Sigma_{o}$ is indeed very complicated. To
simplify the discussion, suppose there is no constraint $H_{2}$
in Fig. 2, i.e., $v(k)\equiv u_{c}(k)$. The fixed points of the system $\Sigma_{o}$ is given by%
\begin{equation}
\left\{ \left( x,x_{d},z\_\right) :x=\frac{3}{2}z_{-},x_{d}=-\frac{1}{2}%
z_{-},\left| z_{-}\right| \leq 2\delta \right\}
\label{fixed_points}
\end{equation}%
Define

\[
\mathbf{x}:=\left[
\begin{array}{ccc}
x & x_{d} & z\_%
\end{array}%
\right] ^{^{\prime }},
\]%
\[
T:=\left[
\begin{array}{ccc}
1 & 0 & -\frac{3}{2} \\
0 & 1 & \frac{1}{2} \\
0 & 0 & 1%
\end{array}%
\right] ,
\]%
\[
\mathbf{\tilde{x}}:=\left[
\begin{array}{ccc}
\tilde{x} & \tilde{x}_{d} & \tilde{z}\_%
\end{array}%
\right] ^{^{\prime }}=T\mathbf{x}.
\]%
Then the system under new coordinates is%
\[
\Sigma _{n1}:\mathbf{\tilde{x}}\left( k+1\right) =\left[
\begin{array}{ccc}
\frac{1}{2} & 3 & -\frac{3}{4} \\
-1 & -2 & -\frac{1}{2} \\
1 & 0 & \frac{3}{2}%
\end{array}%
\right] \mathbf{\tilde{x}}\left( k\right)
\]%
under%
\begin{equation}
\left| \tilde{x}-\tilde{z}\_\right| >\delta ,  \label{sub1_b}
\end{equation}%
and%
\[
\Sigma _{n2}:\mathbf{\tilde{x}}\left( k+1\right) =\left[
\begin{array}{ccc}
2 & 3 & 0 \\
0 & -2 & 0 \\
0 & 0 & 1%
\end{array}%
\right] \mathbf{\tilde{x}}\left( k\right)
\]%
under%
\begin{equation}
\left| \tilde{x}-\tilde{z}\_\right| \leq \delta .  \label{sub2_b}
\end{equation}%
For convenience, we denote this system by $\Sigma _{n}$. It is
easy to see
that the fixed points of $\Sigma _{n}$ are%
\begin{equation}
\left\{ \left( 0,0,\tilde{z}\_\right) :\left| \tilde{z}_{-}\right|
\leq 2\delta \right\} .  \label{new_fixed_points}
\end{equation}%

Some comments are appropriate here:

\begin{itemize}
\item The subsystem $\Sigma _{n1}$ is a stable system.

\item If Eq. (\ref{sub2_b}) is satisfied, a trajectory (governed by $\Sigma
_{n2}$) will move on a surface
\[
\tilde{z}_{-}=\gamma
\]%
for some $\gamma \in \left[ -2\delta ,2\delta \right] $. We call
such a surface $\Omega _{\gamma }$. The point $\left( 0,0,\gamma
\right) $ is the origin of $\Sigma _{n2}$ on $\Omega _{\gamma }$.
Furthermore, the line
\begin{equation}
\Gamma _{s,\gamma }:\tilde{x}=0,\quad \tilde{z}_{-}=\gamma
\label{stable_manifold}
\end{equation}%
is the stable manifold of $\Sigma _{n2}$ and similarly, the line%
\begin{equation}
\Gamma _{u,\gamma }:\tilde{x}_{d}=0,\quad \tilde{z}_{-}=\gamma
\label{unstable_manifold}
\end{equation}%
is the unstable manifold of $\Sigma _{n2}$.
\end{itemize}

Suppose a trajectory $\Gamma $ of the system $\Sigma _{n}$ starts
from a point $p$ and is governed by $\Sigma _{n2}$, if $p\in
\Gamma _{u,\gamma }$ (or in general $p\notin \Gamma _{s,\gamma }$)
on some surface $\Omega _{\gamma }$, then the trajectory will
contract along $\tilde{x}_{d}-$axis and stretch along
$\tilde{x}-$axis. Due to the Eq. (\ref{sub2_b}), after some time,
$\Gamma $ will move according to the stable subsystem
$\Sigma_{n1}$. At this moment, $\Gamma $ will leave the surface $\Omega _{\gamma }$%
, and move toward the origin $\left( 0,0,0\right) $. Due to the
Eq. (\ref{sub1_b}), after some time, it will move again on some surface $%
\Omega _{\gamma ^{^{\prime }}}$ for some $\gamma ^{^{\prime }}\in
\left[ -2\delta ,2\delta \right] $. If it is not exactly on the
line $\Gamma_{s,\gamma ^{^{\prime }}}$, it will once more contract along $\tilde{x}_{d}-$%
axis and stretch along $\tilde{x}-$axis and repeat the above
behavior. So normally a trajectory  never settles done, indicating
its intriguing behavior.

\subsection{Chaotic control?}
The complex dynamical behavior of the system in Fig. 2 has been
studied in detail in the foregoing sections, compared to the
standard control scheme such as that in Fig. 2, whose dynamics can
only be either converging to the origin, or being periodic or
unbounded trajectories, the scheme adopted in Fig. 2 provides much
more dynamical properties. Of course this means that a control
engineer has more flexibility at his/her disposal. This is
particularly attracting from the viewpoint of multi-purpose
control. We believe this is the main merit this control scheme can
provide. In this subsection, we will study the following problem:
Given a control performance specification, can we achieve it by
possibly adjusting the system parameters? We will discuss two
control specifications:
\begin{description}
    \item[(1)] The system has one unique fixed point.
    \item[(2)] A periodic orbit is desirable.
\end{description}
For item (1), without loss of generality, assume that the
desirable unique fixed point is the origin. If the parameter $a$
$\ $in the system in Eq. (\ref{exap1}) satisfies $\left| a\right|
<1$, then we can achieve asymptotic stability with respect to the
origin by adjusting the nonlinear block $H_{1}$, though the system
itself has no such property. According to Fig. 9, \ we need merely
to let the value $v\left( k-1\right) $ stored in $H_{1}$ be $0$
when $\left| x\left( k\right) \right| <\delta $ (This feature is
illustrated in Figs. 10-11). Then the trajectory will move along
the $x-$axis toward the origin, i.e., the asymptotic stability of
the origin is achieved. If the parameter $a$ $\ $in the system in
Eq. (\ref{exap1}) satisfies $\left| a\right| \geq 1$, we can not
expect asymptotic stability of the origin because it itself is
unstable. However, we can keep the trajectory arbitrarily close to
the origin at large time instants,
by adopting the following scheme: Suppose it is desirable to keep the trajectory within the distance $%
\epsilon $ around the origin, then choose $\delta $ small enough so that $%
x\left( k^{\ast }\right) $ satisfies $\left| x\left( k^{\ast }\right)
\right| <\epsilon /$ $\left| a\right| ^{2}$ at some time instant $k^{\ast }$%
(this can be realized, see Fig. 12). Next let $v\left( k^{\ast
}-1\right) =0$ when $\left| x\left( k^{\ast }\right) \right| <\epsilon /$ $\left| a\right| $%
. If $\left| a\right| =1$, then the trajectory will stay at
$\left( 0,x\left( k^{\ast }\right) \right) $ forever. The goal is
achieved. On the other hand, assume $a>1$. If $x\left( k^{\ast
}\right) >0$, we first let the trajectory move along the $x-$axis
until we get $x\left( k^{\ast }+1\right) <\epsilon /$ $\left|
a\right| $, then choose $v\left( k^{\ast }\right) >0$. In this way
$\left( x\left( k^{\ast }+1\right) ,v\left( k^{\ast }\right)
\right) $ is below the line segment of fixed point (then $\left(
x\left( k^{\ast }+1\right) ,v\left( k^{\ast }\right) \right) $
will move downward at the next step) such that%
\begin{equation*}
x\left( k^{\ast }+2\right) >0,
\end{equation*}%
and
\begin{equation*}
x\left( k^{\ast }+2\right) =ax\left( k^{\ast }+1\right) +bv\left( k^{\ast
}\right) <\epsilon /\left| a\right| ^{2}.
\end{equation*}%
(Note this is guaranteed by the property of the vector field of the system.)
Then let $v\left( k^{\ast }+1\right) =0$, and repeat the above procedure.
Similarly if $x\left( k^{\ast }\right) <0$, all we need to do is to choose suitable $%
v\left( k^{\ast }\right) <0$ such that $\left( x\left( k^{\ast
}+1\right) ,v\left( k^{\ast }\right) \right) $ is above the line
segment of fixed points, then follow the above procedure. In this
way we can keep the trajectory within the distance $\epsilon $ of
the origin. Based on the above analysis, we observe that the
instability of the parameter $a$ poses a difficulty for
implementing our scheme, there are more discussions from the
perspective of higher dimensional systems (e.g. Theorem 6 below).
The foregoing discussion is reminiscent of Proposition 2.2 in
Delchamps [1990], however our scheme is better since the $K_{1}$ in that paper can be $%
\infty $ here. \ Moreover, our algorithm is simpler too.

For the item (2), suppose it is desired that the system operates
on a periodic orbit $\Gamma$ of periodic $T$. If $a=1$, according
to Theorem 4, by suitably choosing $b$, a periodic orbit of period
$T$ can be built. If $a\neq 1$, then following the discussion at
the end of Sec. 2.1.2, it is also possible to construct a periodic
orbit of period $T$. Then the real question is: Can we really find
an initial condition which produce or converge to the desirable
periodic orbit $\Gamma$? If $\Gamma$ is within a strange
attractor, then from almost all initial points, trajectories will
be within an arbitrarily small neighborhood of $\Gamma$ at some
time $k$; this is the property of a stranger attractor. So we can
just pick up such an initial condition, let the system run
automatically first, and apply control similarly to the case in
item (1) when the trajectory is sufficiently close to $\Gamma$,
and keep it remain within a small neighborhood of $\Gamma$.
Therefore the problem boils down to constructing a strange
attractor containing $\Gamma$. This is the problem we are
currently studying. Note that our chaotic system seems different
than many known chaotic systems, which have strange attractors
within which there are periodic orbits of any periods. However in
light of Corollary 2, there may be no periodic orbits at all when
$a$ is rational and $b$ is irrational. This annoying fact may
probably be due to the scheme we are proposing involves
discontinuities. We have already known that there may be a great
variety of dynamics this scheme can produce, which brings more
freedom to a control engineer, and especially suitable for
multi-purpose controller design. However in order to make the
proposed scheme more useful, a thorough study of this scheme has
to be conducted.

We have to acknowledge that the preceding analysis is naive,
nevertheless, it illustrates that by using the trajectories of the
system, i.e., some extra information in addition to system
parameters, we can achieve better control in some sense. For
chaotic control, interested readers may refer to Schuster [1999].
These will be our future research directions. Here we still adhere
to classic control theory.

\subsection{Stability analysis of higher-dimensional systems}

Now we return to our analysis of higher dimensional system in Fig.
2. We will
find a positively invariant set for this system. For simplicity, let $%
D_{d}=0 $. Define
\begin{equation*}
\check{A}:=\left[
\begin{array}{cc}
A & BC_{d} \\
-B_{d}C & A_{d}%
\end{array}%
\right] , ~~~ \check{B}:=\left[
\begin{array}{cc}
B & 0 \\
0 & -B_{d}%
\end{array}%
\right] , ~~~ \tilde{B}:=\check{B}\tilde{C}=\left[
\begin{array}{cc}
0 & BC_{d} \\
-B_{d}C & 0%
\end{array}%
\right] .
\end{equation*}

Since the controller $C$ is stabilizing, the closed-loop system in
Fig. 1 is asymptotically stable. Then there exists a Lyapunov
function $v(\xi (k))=$ $\xi ^{^{\prime }}(k)P\xi (k)$ with
$P=\left[
\begin{array}{cc}
P_{1} & P_{2} \\
P_{2}^{^{\prime }} & P_{3}%
\end{array}
\right] >0$ such that
\begin{eqnarray*}
\triangle v(\xi (k)) &=&\xi ^{^{\prime }}(k+1)P\xi (k+1)-\xi ^{^{\prime
}}(k)P\xi (k) \\
&=&\xi ^{^{\prime }}(k)\left( \check{A}^{^{\prime }}P\check{A}-P\right) \xi
(k) \\
&=&-\left\| \xi (k)\right\| _{2}^{2}\mbox{~~for all~~}\xi (k).
\end{eqnarray*}
Correspondingly, define $v_{c}(\eta (k))=$ $\eta ^{^{\prime }}(k)P\eta (k)$,
then
\begin{eqnarray*}
\triangle v_{c}(\eta (k)) &=&\eta ^{^{\prime }}(k+1)P\eta (k+1)-\eta
^{^{\prime }}(k)P\eta (k) \\
&=&\eta ^{^{\prime }}(k)\left( \check{A}^{^{\prime }}P\check{A}-P\right)
\eta (k)+2\eta ^{^{\prime }}(k)\check{A}^{^{\prime }}P\check{B}\left( \left[
\begin{array}{c}
H_{1}\left( u_{c}\left( k\right) ,v(k-1)\right) \\
H_{2}\left( y_{c}\left( k\right) ,z(k-1)\right)%
\end{array}
\right] -\left[
\begin{array}{c}
u_{c}\left( k\right) \\
y_{c}\left( k\right)%
\end{array}
\right] \right) \\
&&+\left( \left[
\begin{array}{c}
H_{1}\left( u_{c}\left( k\right) ,v(k-1)\right) \\
H_{2}\left( y_{c}\left( k\right) ,z(k-1)\right)%
\end{array}
\right] -\left[
\begin{array}{c}
u_{c}\left( k\right) \\
y_{c}\left( k\right)%
\end{array}
\right] \right) ^{\prime }\check{B}^{\prime }\centerdot \\
&&~~~P\check{B}\left( \left[
\begin{array}{c}
H_{1}\left( u_{c}\left( k\right) ,v(k-1)\right) \\
H_{2}\left( y_{c}\left( k\right) ,z(k-1)\right)%
\end{array}
\right] -\left[
\begin{array}{c}
u_{c}\left( k\right) \\
y_{c}\left( k\right)%
\end{array}
\right] \right) \\
&\leq &-\left\| \eta (k)\right\| _{2}^{2}+2\left\| \eta (k)\right\|
_{2}\cdot \left\| \check{A}^{^{\prime }}P\check{B}\right\| _{\infty }\cdot
\gamma \cdot \bar{\delta}+\left( \gamma \cdot \bar{\delta}\right) ^{2}\cdot
\left\| \check{B}^{^{\prime }}P\check{B}\right\| _{\infty },
\end{eqnarray*}
where the positive constant $\gamma =\sqrt{m+p}$ and $\bar{\delta}=\max
\left\{ \delta _{1},\delta _{2}\right\} $. Hence $\triangle v_{c}(\eta
(k))<0 $ if
\begin{equation*}
\left\| \eta (k)\right\| _{2}>\gamma \cdot \bar{\delta}\left\| \check{A}%
^{^{\prime }}P\check{B}\right\| _{\infty }+\gamma \cdot \bar{\delta}\sqrt{%
\left\| \check{A}^{^{\prime }}P\check{B}\right\| _{\infty }^{2}+\left\|
\check{B}^{^{\prime }}P\check{B}\right\| _{\infty }}.
\end{equation*}
For convenience, define
\begin{eqnarray*}
r_{1} &:=&\gamma \cdot \bar{\delta}\left\| \check{A}^{^{\prime }}P\check{B}%
\right\| _{\infty }+\gamma \cdot \bar{\delta}\sqrt{\left\| \check{A}%
^{^{\prime }}P\check{B}\right\| _{\infty }^{2}+\left\| \check{B}^{^{\prime
}}P\check{B}\right\| _{\infty }}, \\
r_{2} &:=&\left\| \check{A}\right\| _{\infty }r_{1}+\left\| \check{B}%
\right\| _{\infty }\bar{\delta},
\end{eqnarray*}
then we have

\begin{theorem}
The set $\Omega $ defined by
\begin{equation*}
\Omega :=\left\{ \eta \left| \eta ^{^{\prime }}(k)P\eta (k)\leq \max \left\{
\bar{\sigma}\left( P\right) r_{1}^{2},\bar{\sigma}\left( P\right)
r_{2}^{2}\right\} \right. \right\}
\end{equation*}
is a positively invariant set, where $\bar{\sigma}\left( P\right) $ is the
largest singular value of $P$.
\end{theorem}

\noindent \textbf{Proof:} We need only to show that for each $\eta
\left( 0\right) \in \Omega $, $\eta \left( k\right) \in \Omega $
for all $k\geq 1$. Suppose for some integer $k_{0}>0$, we have
$\left\| \eta (k_{0})\right\|
_{2}\leq r_{1}$, and $\left\| \eta (k_{0}+1)\right\| _{2}>r_{1}$, then $%
\triangle v_{c}(\eta (k_{0}+1))<0$, which means $\eta ^{^{\prime
}}(k_{0}+2)P\eta (k_{0}+2)<\eta ^{^{\prime }}(k_{0}+1)P\eta
(k_{0}+1).$ Furthermore, the trajectory will eventually fall into
the set $\left\{ \eta \left| \eta ^{^{\prime }}(k)P\eta (k)\leq
\bar{\sigma}\left( P\right) r_{1}^{2}\right. \right\} $. Therefore
it suffices to show $\eta (k_{0}+1)\in \Omega $. Since
\begin{equation*}
\left\| \eta (k_{0}+1)\right\| _{2}\leq \left\| \check{A}\right\| _{\infty
}r_{1}+\left\| \check{B}\right\| _{\infty }\bar{\delta},
\end{equation*}%
one has
\begin{equation*}
\eta ^{^{\prime }}(k_{0}+1)P\eta (k_{0}+1)\leq \bar{\sigma}\left( P\right)
r_{2}^{2},
\end{equation*}%
which gives $\eta (k_{0}+1)\in \Omega $. $\hfill $ $\blacksquare $

\bigskip The preceding result ascertains the existence of a positively invariant set for the system in
Fig. 2, the system behavior insider this invariant set may be very complex.
The next result gives an upper bound for all equilibria of the system in Eq. (%
\ref{switch}).

Defining
\begin{equation}
\Phi :=-\left[
\begin{array}{cc}
I & -C_{d}\left( I-A_{d}\right) ^{-1}B_{d}+D_{d} \\
-C\left( I-A\right) ^{-1}B & I%
\end{array}
\right] ,  \label{thi}
\end{equation}
then we have:

\begin{corollary}
For the system in Eq. (\ref{switch}), supposing both $G$ and $C$
are stable, if the matrix \newline $\left( C_{d}\left(
I-A_{d}\right) ^{-1}B_{d}-D_{d}\right) C\left(
I-A\right) ^{-1}B$ has no eigenvalue at $(-1,0)$, then $\left\| \left( I-%
\tilde{A}\right) ^{-1}\check{B}\right\| _{1}\cdot \left\| \Phi ^{-1}\right\|
_{1}\bar{\delta}$ is an upper bound for all equilibria of this system.
\end{corollary}

\noindent \textbf{Proof: } Suppose $\bar{x}$ is an equilibrium of
the system in Eq. (\ref{switch}), then there are an integer $K>0$
and some vector $\varpi $ such that
\begin{equation}
\eta \left( k+1\right) =\tilde{A}\eta \left( k\right) +\left[
\begin{array}{cc}
B & 0 \\
0 & B_{d}%
\end{array}
\right] \varpi  \label{no-update}
\end{equation}
for all $k>K$. Letting $k\rightarrow \infty $, we get
\begin{equation*}
\bar{x}=\tilde{A}\bar{x}+\left[
\begin{array}{cc}
B & 0 \\
0 & B_{d}%
\end{array}
\right] \varpi ,
\end{equation*}
then
\begin{equation*}
\bar{x}=\left( I-\tilde{A}\right) ^{-1}\left[
\begin{array}{cc}
B & 0 \\
0 & B_{d}%
\end{array}
\right] \varpi ,
\end{equation*}
and
\begin{equation*}
\left\| \tilde{C}\bar{x}+\left[
\begin{array}{cc}
0 & D_{d} \\
0 & 0%
\end{array}
\right] \varpi -\varpi \right\| _{\infty }=\left\| -\Phi \varpi \right\|
_{\infty }.
\end{equation*}

Because the matrix $\left( C_{d}\left( I-A_{d}\right)
^{-1}B_{d}-D_{d}\right) C\left( I-A\right) ^{-1}B$ has no eigenvalue at $(-1,0)$%
, $\Phi $ is invertible. Furthermore, since $\left\| \tilde{C}\bar{x}+\left[
\begin{array}{cc}
0 & D_{d} \\
0 & 0%
\end{array}%
\right] \varpi -\varpi \right\| _{\infty }\leq \bar{\delta}$,
$\left\| \varpi \right\| _{\infty }\leq \left\| \Phi ^{-1}\right\|
_{1}\left\| \Phi \varpi \right\| _{\infty }\leq \bar{\delta}$, we
have
\begin{equation}
\left\| \bar{x}\right\| _{\infty }\leq \left\| \left( I-\tilde{A}\right)
^{-1}\check{B}\right\| _{1}\cdot \left\| \Phi ^{-1}\right\| _{1}\bar{\delta}.
\label{equilibria}
\end{equation}%
Because $\bar{x}$ is arbitrarily chosen, the result follows. $\hfill $ $%
\blacksquare $

\bigskip In particular, assume we have a scalar system with a static state
feedback:
\begin{eqnarray}
x\left( k+1\right) &=&ax\left( k\right) +bv\left( k\right) ,  \notag \\
u\left( k\right) &=&-fx\left( k\right) ,  \label{scalarcase} \\
v\left( k\right) &=&H_{1}\left( u\left( k\right) ,v(k-1)\right) ,  \notag
\end{eqnarray}
where $\left| a-bf\right| <1$. Then following the above procedure, $\left|
\bar{x}\right| \leq \left( \delta \left| bf\right| \right) /\left( 1-\left|
a-bf\right| \right) $ where $\bar{x}$ can be any equilibrium.

An upper bound has been found for all equilibria. Will any of
these equilibria be stable if  either $G$ or $C$ in unstable? We
have a result reminiscent of that in Delchamps [1990]

\begin{theorem}
Assume either $G$ or $C$ is unstable, and $\check{A}$ is invertible, then
the set of all initial points $\eta _{0}$ whose closed-loop trajectories
tend to an equilibrium as $k\rightarrow \infty $ has Lebesgue measure zero.
\end{theorem}

\noindent \textbf{Proof: } Denote this set by $U$. Let $E^{s}$ be
the generalized stable eigenspace of Eq. (\ref{switch}), Then the
Lebesgue measure of $E^{s}$ is zero since Eq. (\ref{switch}) is
unstable. Suppose $\eta \left( 0\right) \in U$, following the
process in the proof of Corollary 1, there exist $K>0$ and some
vector $\varpi $ such that
\begin{equation}
\eta \left( K\right) =\left[
\begin{array}{cc}
A-BD_{d}C & BC_{d} \\
-B_{d}C & A_{d}%
\end{array}%
\right] \varpi ,  \label{unstab1}
\end{equation}%
and Eq. (\ref{no-update}) holds for all $k>K$. Since $\tilde{A}$ is unstable, $%
\eta \left( k\right) \in E^{s}$ for all $k\geq K$. Furthermore,
the invertibility of $\check{A}$ implies that $\varpi $ is
uniquely determined by $\eta \left( K\right) $. Due to the
uniqueness of the state trajectory the system in Eq.
(\ref{switch}), note also that this system is essentially a system
with unit time delay, the trajectory starting from $\left( \eta
\left( -1\right) =0,\eta \left( 0\right) \right) $ is identical to
that starting
from $\left( \varpi ,\eta \left( K\right) \right) $. Define a mapping $%
\digamma $ as
\begin{equation}
\begin{array}{ll}
\digamma : & U\rightarrow E^{s}, \\
& \eta \left( 0\right) \longmapsto \eta \left( K\right) ,%
\end{array}
\label{mapping}
\end{equation}%
where $\eta \left( 0\right) $ and $\eta \left( K\right) $ satisfying Eqs. (\ref%
{unstab1}) and (\ref{no-update}), then $\digamma $ is injective.
Therefore the Lebesgue measure of $U$ is zero. $\hfill $
$\blacksquare $

\section{An Example}
 In this section, one example will be used to illustrate
the effectiveness of the scheme proposed in this paper. In this
example, the networked control system consists of two subsystems,
 (each composed of a system and its controller), the outputs of the
controlled systems will be sent respectively to controllers via a
network. For the ease of notation, we denote the two systems,
their controllers and their outputs by $G_{1}$, $G_{2}$, $C_{1}$,
$C_{2}$, $y_{1}$ and $y_{2}$ respectively. Here two transmission
methods will be compared: one is just letting the outputs
transmitted sequentially, i.e., the communication order is $\left[
y_{1}(0),y_{2}(0),y_{1}(1),y_{2}(1),\cdots \right] $. Another
method is adding the nonlinear constraint $H_{2}$ to the subsystem
composed of $G_{1}$ and $C_{1}$, if the difference between the two
adjacent signals are greater than $\delta _{2}=0.01$, then this
subsystem gets access to the network; otherwise the other gets
access. Here, we will compare the tracking errors produced under
these two schemes respectively. For convenience, we call the first
method the \textit{regular static scheduler} and the second the
\textit{modified static scheduler}.

The controlled system $G_{1}$ is:
\begin{eqnarray*}
x_{1}\left( k+1\right) &=&\left[
\begin{array}{cccc}
1.0017 & 0.1000 & 0.0250 & 0.0009 \\
0.0500 & 1.0000 & 0.5000 & 0.0259 \\
0.2000 & -0.0003 & 1.0000 & 0.1052 \\
-0.0034 & -0.2103 & -0.0517 & 1.1034%
\end{array}
\right] x_{1}\left( k\right) \\
&&+\left[
\begin{array}{c}
0.0050 \\
0.0991 \\
-0.0052 \\
-0.1155%
\end{array}
\right] w\left( k\right) +\left[
\begin{array}{cc}
-0.0050 & -0.0000 \\
-0.1000 & -0.0001 \\
0.0000 & -0.0005 \\
0.0103 & -0.0105%
\end{array}
\right] u_{1}\left( k\right) , \\
z_{1}\left( k\right) &=&\left[
\begin{array}{cccc}
1 & 0 & 0 & 0 \\
1 & 0 & -1 & 0%
\end{array}
\right] x_{1}\left( k\right) +\left[
\begin{array}{c}
-1 \\
0%
\end{array}
\right] w\left( k\right) , \\
y_{1}\left( k\right) &=&\left[
\begin{array}{cccc}
1 & 0 & 0 & 0 \\
0 & 0 & 1 & 0%
\end{array}
\right] x_{1}\left( k\right) ,
\end{eqnarray*}
and $G_{2}$ is:
\begin{eqnarray*}
x_{2}\left( k+1\right) &=&\left[
\begin{array}{cccc}
1.0000 & 0.0100 & 0.0002 & 0.0000 \\
0.0005 & 1.0000 & 0.0500 & 0.0003 \\
0.0200 & -0.0000 & 1.0000 & 0.0101 \\
-0.0000 & -0.0201 & -0.0005 & 1.0100%
\end{array}
\right] x_{2}\left( k\right) \\
&&+\left[
\begin{array}{c}
0.0000 \\
0.0100 \\
-0.0001 \\
-0.0102%
\end{array}
\right] w\left( k\right) +\left[
\begin{array}{cc}
-0.0000 & -0.0000 \\
-0.0100 & -0.0000 \\
0.0000 & -0.0000 \\
0.0001 & -0.0010%
\end{array}
\right] u_{2}\left( k\right) , \\
z_{2}\left( k\right) &=&\left[
\begin{array}{cccc}
1 & 0 & 0 & 0 \\
1 & 0 & -1 & 0%
\end{array}
\right] x_{2}\left( k\right) +\left[
\begin{array}{c}
-1 \\
0%
\end{array}
\right] w\left( k\right) , \\
y_{2}\left( k\right) &=&\left[
\begin{array}{cccc}
1 & 0 & 0 & 0 \\
0 & 0 & 1 & 0%
\end{array}
\right] x_{2}\left( k\right) ,
\end{eqnarray*}
where $w$ is a unit step. $z_{1}$ and $z_{2}$ are tracking errors.
Controllers $C_{1}$ and $C_{2}$ can be obtained using the
technique in Chen \& Francis [1995]. Denote the first element of
$y_{1}$ by $y_{11}$ and that of $y_{2}$ by $y_{21}$; the second
element of $y_{1}$ by $y_{12}$ and that of $y_{2}$ by $y_{22}$,
then the subsystem with variables
$x_{1},x_{2},z_{1},z_{2},y_{11},y_{21}$ is $G_{1}$ controlled by
$C_{1}$ and the subsystem with variables
$x_{1},x_{2},z_{1},z_{2},y_{12},y_{22}$ is $G_{2}$ controlled by
$C_{2}$. The simulation results are in Figs. 18--19.
\begin{figure}[tbh]
\epsfxsize=4in
\par
\epsfclipon
\par
\centerline{\epsffile{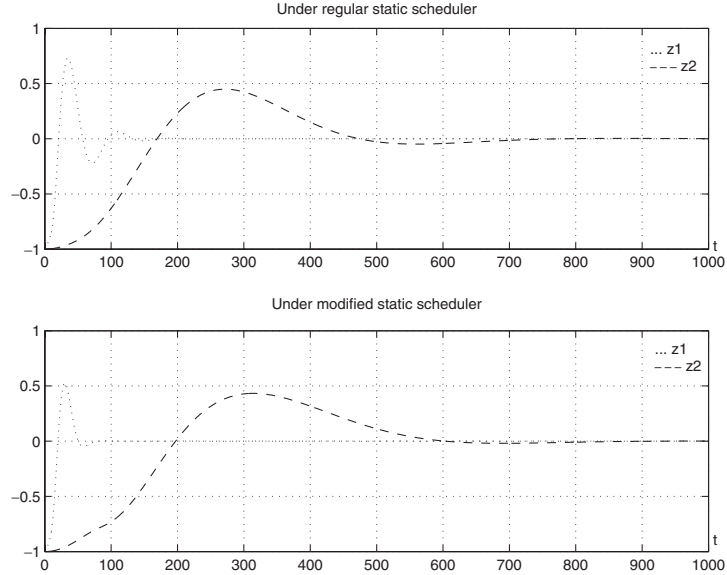}} \caption{The first elements
of $z_{1}$ and $z_{2}$}
\end{figure}

\begin{figure}[tbh]
\epsfxsize=4in
\par
\epsfclipon
\par
\centerline{\epsffile{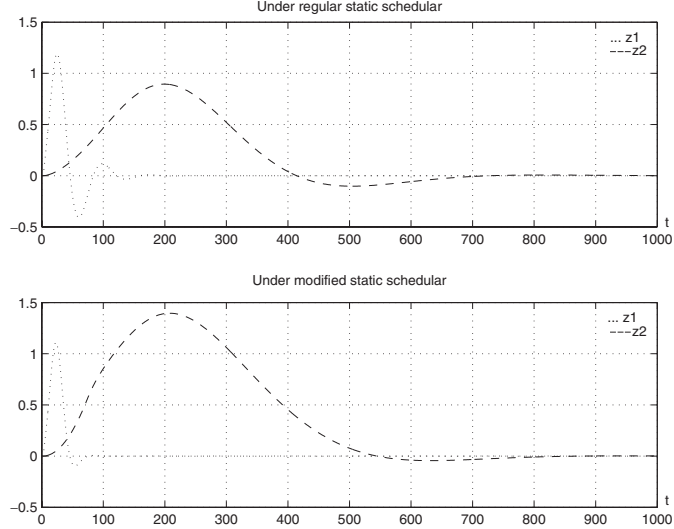}} \caption{The second elements
of $z_{1}$ and $z_{2}$}
\end{figure}

From these two figures, one finds that the tracking errors
approach zero faster under the modified static schedular than
under the regular one. Note that both systems are unstable. If one
of the two systems is stable, one can expect better convergence
rate. In essence, our scheme is based on the following principle:
Allocate access to the network to the systems with faster dynamics
first, then take care of the systems of slower dynamics. In this
way, we hope we can improve system performance. Interestingly, a
similar idea is explored in Hristu \& Morgansen [1999].

\section{Conclusion}

In this paper, a new networked control technique is proposed and
its effectiveness is illustrated via simulations. The complicated
dynamics of this type of systems is studied both numerically and
theoretically. A simulation shows that the scheme proposed here
has possible application in networked control systems. There are
several problems guiding our further research: 1) Continuity of
state trajectories with respect to the initial points under space
partition induced by the discontinuities of the system. 2) How to
find a precise characterization of the attracting set for our
system, and is it topologically transitive (i.e., is it a chaotic
attractor)? Topological transivity, an indispensable feature of a
chaotic attractor, is closely related to ergodicity of a map. As
discussed in Sec. 2.1.3, the proof of topological transitivity or
ergodicity is difficult for our system from the point of view of
measure theory due to the singularity of the map and its violation
of conditions in Lasota \& Yorke [1973]. However, this
investigation is unavoidable should one want to find the chaotic
attractor inherited in the system studied. 3) For different system
parameters, different aperiodic orbits can be obtained, what are
the differences among these orbits? In particular, given two
aperiodic orbits, one generated from a system having no periodic
orbits and the other generated by a system having periodic orbits,
is there any essential difference between them? 4) In Sec. 2.1.2,
periodic orbits are constructed for some originally stable
($|a|<1$) and originally unstable ($|a|>1$) systems. However given
a system, how to determine if there are periodic orbits, and if
so, how to find all of them is still an unsolved problem.
 5) How to effectively design controllers based on
chaotic control? Obviously the solution of this problem depends on
the forgoing ones. 6) How to incorporate properly the scheme
proposed in this paper into the framework of networked control
systems? The simulation in Sec. 3 is naive, more research is
required here to make the proposed scheme practical.

\section{Acknowledgement}
The first author is grateful to discussions with Dr. Michael Li.
This work was partially supported by NSERC. The authors are also
grateful to the anonymous reviewers and the Editor for their
resourceful comments and constructive suggestions.

\end{document}